\documentclass[preprints,review,accept,moreauthors,pdftex]{mdpi} 

\makeatletter							
\renewcommand{\maketag@@@}[1]{\hbox{\m@th\normalsize\normalfont#1}}%
\makeatother							

\firstpage{1} 
\pubvolume{1}
\issuenum{1}
\articlenumber{0}
\pubyear{2021}
\copyrightyear{2021}
\externaleditor{} 
\datereceived{} 
\dateaccepted{} 
\datepublished{} 
\hreflink{https://doi.org/} 

\pdfoutput=1

\usepackage{mathtools}
\usepackage[thinc]{esdiff}

\Title{Classical and Quantum $f(R)$ Cosmology: The Big Rip, the Little Rip and the Little Sibling of the Big Rip}

\TitleCitation{Classical and Quantum $f(R)$ Cosmology: The Big Rip, the Little Rip and the Little Sibling of the Big Rip}

\Author{Teodor Borislavov Vasilev $^{1,}$*\orcidA{}, Mariam Bouhmadi-L\'opez $^{2,3}$\orcidB{} and Prado Mart\'in-Moruno $^{1}$\orcidC{}}

\AuthorNames{Teodor Borislavov Vasilev, Mariam Bouhmadi-L\'opez, Prado Mart\'in-Moruno}

\AuthorCitation{Borislavov Vasilev, T.; Bouhmadi-L\'opez, M.; Mart\'in-Moruno, P.}

\address{%
	$^{1}$ \quad Departamento de F\'isica Te\'orica and IPARCOS,
	Universidad Complutense de Madrid,
	28040 Madrid, Spain; pradomm@ucm.es\\
	$^{2}$ \quad IKERBASQUE, Basque Foundation for Science, 48011 Bilbao, Spain; mariam.bouhmadi@ehu.eus\\
	$^{3}$ \quad
	Department of Theoretical Physics, University of the Basque Country,
	UPV/EHU, P.O. Box 644, 48080~Bilbao,~Spain}

\corres{Correspondence: teodorbo@ucm.es}

\abstract{The big rip, the little rip and the little sibling of the big rip are cosmological doomsdays predicted by some phantom dark-energy models that could describe the future evolution of our universe. When the universe evolves towards either of these future cosmic events, all bounded structures and, ultimately, space–time itself are ripped apart. Nevertheless, it is commonly believed that quantum gravity effects may smooth or even avoid these classically predicted singularities. In this review, we discuss the classical and quantum occurrence of these riplike events in the scheme of  metric $f(R)$ theories of gravity. The quantum analysis is performed in the framework of $f(R)$ quantum geometrodynamics. In this context, we analyze the fulfilment of the DeWitt criterion for the avoidance of these singular fates. This review contains as well new unpublished work (the analysis of the equation of state for the phantom fluid and a new quantum treatment of the big rip and the little sibling of the big rip events).}


\begin{document}

	\section{Introduction}
	
	The discovery that our universe is currently undergoing an accelerated expansion phase has supposed an inflexion point in our understanding of the physics of the cosmos. Even though this acceleration was first noticed more than 20 years \mbox{ago \cite{Riess,Perlmutter}}, its mysteries are still not yet fully unraveled. On the contrary, understanding the mechanism involved in this phase represents one of the greatest milestones in modern cosmology. This behavior is so challenging because no form of matter we know from our ordinary experience can actually produce this phenomenon. In general relativity (GR), this phase is attributed to the existence of an exotic form of energy with a negative pressure that causes the repulsion of the matter content and, thus, pushes the universe into expansion. This exotic content is dubbed ``dark energy'' (DE). The simplest DE model, and still the one which best fits the observational data, is the standard $\Lambda$CDM theory, where DE plays the role of a cosmological constant. Nevertheless, this model  contains various theoretical and observational unresolved puzzles. To mention some of them: the nature of dark matter (DM) and DE \cite{Weinberg:2000yb,Peebles:2002gy,Padmanabhan:2002ji,Sahnidmde,Copeland:2006wr}, the coincidence problem \cite{Velten:2014nra}, the cosmological constant problem \cite{LambdaProblem} and the new tensions in certain cosmological parameters. (For a review of the topic see, for instances, references \cite{Bull:2015stt,Perivolaropoulos:2021jda} and references therein.) Therefore, one may expect the $\Lambda$CMD paradigm to be just a useful effective model. Hence, a great number of alternative models for DE describing the acceleration phase have been proposed. Some models are phantom DE \cite{Starobinsky:1999yw,PhantomDE}, tachyonic matter \cite{TachyonGibbons:2002md,TachyonPadmanabhan:2002cp}, Chaplygin gas \cite{ChaplyginKamenshchik:2001cp,ChaplyginBento:2002ps}, holographic \mbox{DE \cite{HoloLi:2004rb}} and scalar fields in the form of quintessence \cite{Caldwell,Tsujikawa} and k-essence \cite{kessenceChiba:1999ka}, among other examples.
	Alternatively, the current cosmological expansion could be described not by  the inclusion of new exotic content but with suitable modifications to the underlying theories of gravity.  In that sense, modified theories of gravity provide an interesting framework for cosmologists. Some examples of this approach are Horndeski \mbox{theories \cite{Horndeski}} (see also, e.g., reference \cite{Kase:2018aps}), Gauss-Bonnet gravity \cite{GB1Dehghani:2004cf,GB2Nojiri:2005jg}, $f(R)$ theories of gravity \cite{DEinfRNojiri} (see also references \cite{Nojiri:2004bi, Allemandi:2005qs,Capozziello:2011et}), $f(R,\mathcal{T})$ gravity \cite{DEinfRTHarko:2011kv}, where $\mathcal{T}$ stands for the trace of the energy momentum tensor,  $f(T)$ modified teleparallel gravity \cite{DEinfTBengochea:2008gz}, being $T$ the torsion scalar, and modified symmetric teleparallel $f(Q)$ theories of gravity \cite{Jimenez:2019ovq}, where $Q$ denotes the non-metricity scalar. (See, for example, references \cite{Saridakis:2021lqd,Huterer:2017buf,reviewDE, Motta:2021hvl} and references therein for a review of the state of the art.) 
	
	From a practical perspective, whatever the origin of DE may be, it can be described effectively by its equation of state (EoS) parameter. That is the ratio  $w$ between the pressure and energy density of the DE, or the effective pressure and energy density in the case of modified theories of gravity.  Hence, observational constraints on $w$ are crucial to unveil possible hints on the true nature of this exotic content. In this fashion, the cosmological data currently available constrain $w$ to a very narrow band around $-$1 \cite{Planck1,Planck2}, see \mbox{also \cite{Abbott:2018wog, Lopez-Corredoira:2016pwg, wCDM1,wDCM2,Addison:2013haa,Santos:2015nua,Bonilla:2017ygx}}. Accordingly, the possibility of the expansion of the universe being fueled by phantom-type DE ($w<-1$) is not observationally excluded. Furthermore, it is even suggested by some data \cite{RisalitiLusso} and could help to alleviate the $H_0$ tension (see, for instance, \mbox{references \cite{EDVH01,EDVH02})}. Nevertheless, considering phantom DE may entail some future cosmological singularities. (See reference \cite{BouhmadiLopez:2004me} for counterexample.) All bounded structures in the universe and, ultimately, the fabric of space–time itself might be torn apart at a final big rip (BR) \mbox{singularity \cite{BR}}. Moreover, the universe could reach this singularity in a finite time from present epoch, resulting in a moment at which the size of the observable universe, the Hubble rate and its cosmic time derivative would diverge. On the other hand, for a phantom DE model with a $w$ parameter  converging sufficiently rapidly to the cosmological constant value ($w=-1$), the occurrence of a future singularity may be infinitely delayed in time \cite{LR}. In that case, the singularity is called an abrupt cosmic event. This is precisely the case of the little rip (LR) abrupt event \cite{EOSalpha2table,BouhmadiLopez:2005gk} (see also \cite{LR,LR2}), where the scale factor, the Hubble rate and its cosmic time derivative explode at the infinite asymptotic future.
	Therefore, this abrupt event can be understood as a BR singularity that has been postponed indefinitely. Another characteristic abrupt event of phantom DE models is the little sibling of the big rip (LSBR). In this scenario, the size of the observable universe and the expansion rate grow infinitely but the cosmic time derivative of the Hubble parameter converges to a constant \mbox{value \cite{LSBR}}. It should be noted, however, that even though these abrupt events take place at the infinite distant future, bounded structures are destroyed in a finite time \cite{LR,LSBR}. See a brief summary of the phenomenology of these riplike doomsdays in \mbox{Table \ref{tab1}}. (See also \mbox{references \cite{CosmoConstraints2,CosmoConstraints}} for observational constraints on some phantom DE models leading the expansion towards these events.)

	The BR singularity, and the LR and the LSBR abrupt events are, in fact, intrinsic to phantom DE models. Nevertheless, a dark fluid could also induce the appearance of other cosmic singularities. For example, a phantom-dominated universe could also reach a big freeze singularity (BF) \cite{BF1BouhmadiLopez:2006fu, BF2BouhmadiLopez:2007qb}, where the scale factor reaches a maximum size at which the Hubble rate and its cosmic time derivative diverge. (See also references \cite{sudden1,sudden2,sudden3} for other examples of cosmological singularities  and references \cite{EOSalpha3,Dabrowski:2014fha, SingularityClasification} for a detailed classification of DE singularities in cosmology.) However, it is commonly believed that a consistent quantum description of the universe may prevent the appearance of classical singularities, see references \cite{Dabrowski:2006dd,Kamenshchik:2007zj} (see also \cite{EOSalpha3,SingularityClasification,EOSalpha1,Elizalde:2004mq,Nojiri:2004ip,BouhmadiLopez:2009pu}).

	The quantum fate of classical singularities can be addressed in the setup of quantum cosmology: the application of quantum theory to the universe as a whole.
	Previous works in this framework have shown that the aforementioned phantom riplike cosmological doomsdays, among other singularities such as the big bang, can be avoided as a result of quantum effects emerging as the universe approaches the classical singularity \cite{Dabrowski:2006dd,GRLR,Albarran:2015cda}. However, since the same classical background evolution can be equivalently described in the context of GR or by alternative theories of gravity, it is natural to wonder whether these singularities are still avoided in the quantum realm for a different underlying theory of gravity. In this work, we gather and review the so far published results about the classical and quantum occurrence of these three riplike events when the description of gravity is that provided by  $f(R)$ metric theories of gravity. For that purpose, we shall focus on the quantum cosmology scheme given by the Wheeler-DeWitt (WDW) equation \cite{DeWitt} being adapted for the case of $f(R)$ gravity \cite{Vilenkin}, namely $f(R)$ quantum geometrodynamics. Consequently, we explore here the possibility of avoiding the BR singularity, and the LR and the LSBR abrupt events in $f(R)$ quantum cosmology.
	
	This manuscript is organized as follows. In Section \ref{sec:PhantomDE}, we review some phantom DE models predicting the classical appearance of future singularities and/or abrupt events. We also provide a new classification of the cosmic singularities and abrupt events found there. Moreover, we discuss the viability of those models from the point of view of cosmological observations.
	In Section \ref{sec:fR}, we consider that the classical background evolution found in the previous section can be equivalently described in the framework of metric $f(R)$ theories of gravity, where in the latter case the expansion of the universe will have a purely geometrical origin. For that aim, we briefly discuss a reconstruction method for metric $f(R)$ theories of gravity. Thereafter, in Sections \ref{sec:BRinfR}--\ref{sec:LSBRinfR}, we apply this background reconstruction technique to the specific phantom DE models reviewed in Section \ref{sec:BR_LR_LSBR}. Thus, we present the group of metric $f(R)$ theories of gravity predicting a classical fate \textit{\`a la} BR, LR or LSBR. Section \ref{sec:QC} is entirely devoted to the revision of the fate of  classical singularities in $f(R)$ quantum cosmology. Therefore, we summarize the  $f(R)$ quantum geometrodynamics approach in Section \ref{sec:mWDW}. Then, the resolution of the modified Wheeler-DeWitt equation corresponding to the BR, LR and LSBR doomsdays is addressed in Sections \ref{sec:mWDWBR}--\ref{sec:mWDWLSBR}, respectively. We also discuss there the avoidance of these cosmological catastrophes by means of the DeWitt criterion. Finally, we check the validity of the approximations performed to solve the modified Wheeler–DeWitt equations in Appendices \ref{ap:BR} and \ref{ap:BO}.
	
	\begin{specialtable}[H]
		\caption{Classification of the riplike cosmic events by means of the time of occurrence of the rip $t_{\textup{rip}}$, the scale factor $a$, the Hubble parameter $H$ and its cosmic time derivative $\dot{H}$. Please note that the pseudorip has been included for the sake of completeness, as this corresponds to a mild event (before which all bounded structures are disintegrated) rather than to a curvature singularity.
			\label{tab1}}
		\setlength{\tabcolsep}{9.2mm}
		\begin{tabular}{lcccc}
			\toprule
			& \boldmath{\textbf{$t_{\textup{rip}}$}}	& \boldmath{\textbf{$a$}}&\boldmath{\textbf{$H$}}&\boldmath{\textbf{$\dot{H}$}}\\
			\midrule
			Big rip		& Finite   & $\infty$ & $\infty$ & $\infty$\\
			Little rip	& $\infty$ & $\infty$ & $\infty$ & $\infty$ \\
			LSBR        & $\infty$ & $\infty$ & $\infty$ & Finite\\
			Pseudorip   & $\infty$ & $\infty$ & Finite   & Finite\\
			\bottomrule
		\end{tabular}
	\end{specialtable}

	
	\section{A Phantom Dark-Energy Universe\label{sec:PhantomDE}}
	Throughout this review we consider homogeneous and isotropic cosmological scenarios, which are described by the Friedmann-Lema\^itre-Robertson-Walker (FLRW) metric given by
	\begin{eqnarray}
		ds^2=-dt^2+a(t)^2ds_3^2,
	\end{eqnarray}
	where $t$ stands for the cosmic time, $a$ represents the scale factor, $ds_3^2$ denotes the three-dimensional spatial metric and we have used the geometric unit system $8\pi G=c=1$. Assuming the content of the universe to be described by a perfect fluid with energy density $\rho$ and pressure $p$, the Friedmann and Raychaudhuri equations are
	\begin{align}
		H^2=& \frac13\rho-\frac{k}{a^2},\label{eq:FE}\\
		\dot{H}+H^2=&-\frac12\left(p+\frac{\rho}{3}\right)\label{eq:FE2},
	\end{align}
	respectively, where the dot denotes the derivative with respect to the cosmic time $t$, $H$ stands for the Hubble rate and $k$ represents the spatial curvature (not fixed at this point). Within the standard interpretation of the observational data, the cosmic fluid at present is constituted by three species: radiation, matter (baryonic and dark) and dark energy. The present concentrations of matter and dark energy are roughly $\Omega_{M,0}=0.315$ and $\Omega_{DE,0}=0.685$, respectively, whereas the contribution of radiation to  the total content of the universe is negligible at the present time \cite{Planck1,Planck2}. Therefore, DE is the dominant ingredient today. Moreover, because we aim to study the asymptotic future expansion of the universe, we assume that the DE density increases, remains constant or decreases more slowly than the non-relativistic matter energy density and the spatial curvature term $k/a^2$. Thus, from a practical point of view, we can neglect the contribution of the other cosmic ingredients when studying the future asymptotic behavior of these models. Consequently, hereafter $p$ and $\rho$ denote the pressure and the energy density of the dark fluid. Then, the DE density evolves as 
	\begin{eqnarray}\label{eq:conservationRho}
		\dot{\rho}+3H(p+\rho)=0.
	\end{eqnarray}
	
	To evaluate the future fate of a DE dominated universe in the perfect fluid representation, an equation of state (EoS) for the dark fluid must be provided. We emphasize that presently observational data are compatible with the $w=-1$ value, though there is a significantly overlap with both possibilities $w<-1$ and $w>-1$. On the other hand, from a theoretical point of view, the value $-1$ in the $w$-axis represents a qualitative change in the properties of DE, as well as in the future evolution of the universe. Therefore, the cosmological constant case ($w=-1$) seems to play a special role from both theoretical and observational perspectives. Then, let us begin with a brief review of the phenomenology of the following EoS for the DE content, which is reminiscent of an expansion around a cosmological constant,
	\begin{eqnarray}\label{eq:EoS}
		p=-\rho-A\rho^\alpha,
	\end{eqnarray}	
	being $A$ a positive constant
	\footnote{Since the BR, the LR and the LSBR doomsdays are intrinsic to phantom-like DE models, we limit our discussions to a positive parameter $A$, which leads to $w<-1$. For an analysis of the most general case, we refer the reader to the reference \cite{EOSalpha2table}.}. This equation of state was first introduced in \mbox{references \cite{EOSalpha1,EOSalpha3}} and thoroughly discussed in terms of singularity occurrence in reference \cite{EOSalpha2table}. 
	In this section, of the manuscript, we review the analysis performed in reference \cite{EOSalpha2table} and present a novel alternative classification of the singularities there found by means of the behavior of the scale factor $a$, the Hubble rate $H$ and its cosmic time derivative $\dot{H}$. That is a metric classification such as the one presented in \cite{SingularityClasification}. (See also the characterization given in \cite{EOSalpha3} and the  study on cosmic singularities presented in references \cite{Fernandez-Jambrina:2014sga,Fernandez-Jambrina:2016clh}.)  We argue for the need for such a classification, instead of only addressing the evolution of the scale factor $a$ and the DE density $\rho$, since different cosmic events such as the LR and LSBR cannot be differentiated in the latter picture.
	
	From the continuity equation (\ref{eq:conservationRho}) it follows that the DE density evolves as
	\begin{eqnarray}\label{eq:rho}
		\rho=\rho_0\left[1+\frac{3\left(1-\alpha\right)A}{\rho_0^{1-\alpha}}\ln\left(\frac{a}{a_0}\right)\right]^{\frac{1}{1-\alpha}},
	\end{eqnarray}
	for $\alpha\neq1$, and
	\begin{eqnarray}
		\rho=\rho_0\left(\frac{a}{a_0}\right)^{3A},
	\end{eqnarray}
	for the case of $\alpha=1$, where the subscript ``0'' denotes the current value of the corresponding quantity. In either case, the EoS parameter $w$ reads
	\begin{eqnarray}\label{eq:w}
		w=-1-\frac{A}{\rho_0^{1-\alpha}+3(1-\alpha)A\ln\left(\frac{a}{a_0}\right)}.
	\end{eqnarray}
	Note that for a non-negative parameter $A$, the denominator in the r.h.s. in equation (\ref{eq:w}) is always positive \footnote{This is trivial to check when $\alpha\leq1$. For the case of $\alpha>1$, on the other hand, the expansion of the universe stops at a finite value of the scale factor, namely $a_\textup{max}$, such as  the denominator in equation (\ref{eq:w}) never becomes negative [see equation (\ref{eq:amax})].}, whatever the value of $\alpha$.
	Thus, the DE content modelled by the EoS (\ref{eq:EoS}) exhibits a phantom-like behavior when the parameter $A$ is positive. Please note that this can be also seen directly from equation (\ref{eq:conservationRho}) for an expanding ($H>0$) universe.

	On the other hand, note that $\alpha=1$ and $\alpha=\frac12$ are special values on the $\alpha$-line that delimited qualitatively alike cosmological behaviors \cite{EOSalpha2table}. This can be deduced from  the time dependence of the scale factor obtained from equations (\ref{eq:FE}) and (\ref{eq:rho}). This is \cite{EOSalpha2table}
	\begin{align}\label{eq:a(t)LR}
		\ln \left(\frac{a}{a_0}\right)=
		\frac{2\sqrt{\rho_0}}{3A}\left\lbrace\left[1+\frac{3A}{2\sqrt{\rho_0}}\ln\left(\frac{a_\star}{a_0}\right)\right]\exp\left(\frac{\sqrt{3}A}{2}\left(t-t_\star\right)\right)-1\right\rbrace,
	\end{align}
	for the case of $\alpha=\frac12$, and \cite{EOSalpha2table}
	\begin{eqnarray}\label{eq:a(t)not0.5}
		\ln \left(\frac{a}{a_0}\right)=\frac{1}{3(\alpha-1)\rho_0^{\alpha-1}A}\left\lbrace1- \left[B-\frac{\sqrt{3}}{2}(2\alpha-1)\rho_0^{\alpha-\frac12}A(t-t_\star)\right]^{\frac{2(\alpha-1)}{2\alpha-1}}\right\rbrace,
	\end{eqnarray}
	for $\alpha\neq\frac12$ and $\alpha\neq1$, and
	\begin{eqnarray}\label{eq:a alpha 1}
		\frac{a}{a_0}=\left[\left(\frac{a_\star}{a_0}\right)^{-\frac{3A}{2}}-\frac{\sqrt{3\rho_0}}{2}A\left(t-t_\star\right)\right]^{-\frac{2}{3A}},
	\end{eqnarray}
	for $\alpha=1$. $B$ is a constant defined as
	\begin{eqnarray}
		B\coloneqq\left[1+3(1-\alpha)\rho_0^{\alpha-1}A\ln\left(\frac{a_\star}{a_0}\right)\right]^{\frac{2\alpha-1}{2(\alpha-1)}},
	\end{eqnarray}
	where we have denoted by $t_\star$ some arbitrary (future) moment in the expansion history of the universe from which we can safely neglect the contribution of matter fields and, therefore, assume that DE is the only content of the cosmos. In addition, $a_\star$ represents the corresponding scale factor.
	Therefore, when $\alpha=\frac12$ the scale factor asymptotically approaches a double exponential growth on the cosmic time. However, for $\alpha>\frac12$, it evolves as some function of $t_{\textup{s}}-t$, being $t_{\textup{s}}$ the time of occurrence of the corresponding singularity. Moreover, the expression for $t_{\textup{s}}$ depends on the value of $\alpha$. Hence, for the sake of the discussion, we proceed to summarize the main results found in \cite{EOSalpha2table} for each case. Those are: $\alpha>1$, $\alpha=1$, $\frac12<\alpha<1$, $\alpha=\frac12$ and $\alpha<\frac12$.
	Additionally, to the conclusion presented in~\cite{EOSalpha2table}, we also compute the corresponding $H$ and $\dot{H}$ variables in each case. This allows us to provide a complementary classification of the cosmic events found in \cite{EOSalpha2table} similar to that introduced in reference \cite{SingularityClasification}, see Table \ref{tab:EoSalpha}.
	
	For the case of $\alpha>1$, the universe reaches a maximum size given by
	\begin{eqnarray}\label{eq:amax}
		a_{\textup{max}}\coloneqq a_0 \exp\left(\frac{1}{3(\alpha-1)\rho_0^{\alpha-1}A}\right).
	\end{eqnarray}
	Furthermore, this size is reached in a finite time into the future
	\begin{eqnarray}
		t_{\textup{s}}\coloneqq t_\star +\frac{2B}{\sqrt{3}(2\alpha-1)\rho_0^{\alpha-\frac12}A}.
	\end{eqnarray}
	At that moment, the Hubble rate and its comics time derivative diverge, since
	\begin{align}\label{eq:HBRBF}
		H=&\sqrt{\frac{\rho_0}{3}}\left[\frac{\sqrt{3}}{2}(2\alpha-1)\rho_0^{\alpha-\frac12}A(t_{\textup{s}}-t)\right]^{-\frac{1}{2\alpha-1}},\\
		\dot{H}=&\frac{\rho_0^\alpha A}{2}\left[\frac{\sqrt{3}}{2}(2\alpha-1)\rho_0^{\alpha-\frac12}A(t_{\textup{s}}-t)\right]^{-\frac{2\alpha}{2\alpha-1}}.\label{eq:dotHBRBF}
	\end{align}
	This asymptotic behavior corresponds to the occurrence of a BF singularity \cite{BF1BouhmadiLopez:2006fu,BF2BouhmadiLopez:2007qb}(see also reference \cite{Fernandez-Jambrina:2014sga}). That is a type III singularity in the notation of \cite{EOSalpha3}. 
	
	The case of $\alpha=1$ corresponds to a constant EoS parameter $w=-1-A$. Accordingly, the size of the observable universe becomes infinite at a finite time from present epoch, namely $t_{\textup{rip}}$. This is
	\begin{eqnarray}
		t_{\textup{rip}}\coloneqq t_\star+\frac{2}{\sqrt{3\rho_0}A}\left(\frac{a_\star}{a_0}\right)^{-\frac{3A}{2}},
	\end{eqnarray}
	see equation (\ref{eq:a alpha 1}). Furthermore, given that
	\begin{align}
		H=&\frac{2}{3A\left(t_{\textup{rip}}-t\right)},\label{eq:H alpha 1}\\
		\dot{H}=&\frac{2}{3A\left(t_{\textup{rip}}-t\right)^2}\label{eq:dotH alpha 1},
	\end{align}
	the Hubble rate and its cosmic time derivative also diverge at $t=t_{\textup{rip}}$.
	Therefore, for this value of $\alpha$, the universe evolves towards a classical BR singularity, alike to that first introduced in \cite{Starobinsky:1999yw,PhantomDE,BR}. This corresponds to a type I singularity according to the notation in reference \cite{EOSalpha3}.

	On the other hand, for $\frac12<\alpha<1$, the scale factor diverge at finite cosmic time, see equation (\ref{eq:a(t)not0.5}) where the ratio $2(\alpha-1)/(2\alpha-1)$ is now negative. This makes the $\ln a$  proportional to some power of $1/(t_{rip}-t)$. The time at which the observable universe becomes infinite is
	\begin{eqnarray}
		t_{\textup{rip}}\coloneqq t_\star+\frac{2B}{\sqrt{3}(2\alpha-1)\rho_0^{\alpha-\frac12}A}.
	\end{eqnarray}
	The Hubble rate and its time derivative follows the same relations given in \mbox{equations (\ref{eq:HBRBF})} and (\ref{eq:dotHBRBF}), respectively. Hence, these quantities also diverge along with the scale factor. Therefore, in a finite time into the future, the scale factor, the Hubble rate and $\dot{H}$ explode, whereas the EoS parameter $w$ converges to $-1$ from below. Of course, this implies that the DE density and pressure blow up, as was found in \cite{EOSalpha2table}. This event is qualitatively equivalent to a BR singularity (see the singularities classified as type I in, for example, references \cite{EOSalpha3,Dabrowski:2014fha,SingularityClasification}). Please note that this behavior was also found in reference \cite{Fernandez-Jambrina:2014sga}, where it was dubbed ``grand rip'' (see type $-$1 singularities in reference \cite{Fernandez-Jambrina:2016clh}).
	
	Another possible choice for $\alpha$ corresponds to the interesting case of $\alpha=\frac12$. In this scenario, the scale factor asymptotically evolves as an exponential function of an exponential function on the cosmic time, i.e., $a\approx e^{e^t}$, see equation (\ref{eq:a(t)LR}). Accordingly, the Hubble rate and its cosmic time derivative are
	\begin{align}
		H(t)&=\sqrt{\frac{\rho_0}{3}}\left(1+\frac{3A}{2\sqrt{\rho_0}}\ln \frac{a_\star}{a_0}\right)\exp\left[\frac{\sqrt{3}}{2}A\left(t-t_\star\right)\right],\label{eq:HLR}\\
		\dot{H}(t)&=\frac{A}{2}\sqrt{\rho_0}\left(1+\frac{3A}{2\sqrt{\rho_0}}\ln \frac{a_\star}{a_0}\right)\exp\left[\frac{\sqrt{3}}{2}A\left(t-t_\star\right)\right]\label{eq:dotHLR}.
	\end{align}
	Thus, the scale factor, $H$, and $\dot{H}$ diverge at the infinite asymptotic future. This drives the universe towards a LR abrupt event, see classification in \cite{Dabrowski:2014fha,SingularityClasification}. (See also reference \cite{BouhmadiLopez:2005gk}.) In fact, the EoS (\ref{eq:EoS}) with $\alpha=\frac12$ corresponds to the DE model for which the name ''little rip'' was first given \cite{LR}, even though this cosmological behavior was already known from before \cite{EOSalpha2table} (see also \cite{BouhmadiLopez:2005gk} where the LR was found in brane cosmology and before that in some modified theories of gravity \cite{Ruzmaikina}).
	
	Finally, for the case of $\alpha<\frac12$, the scale factor obeys the relation given in equation \break (\ref{eq:a(t)not0.5}). However, since now the ratio $2(\alpha-1)/(2\alpha-1)$ is positive, then the equation for $\ln a$ reduces to a certain polynomial on the cosmic time. This makes the expansion of the observable universe to last indefinitely. Hence, no  finite time singularities are present in this case. The Hubble rate and its cosmic time derivative read
	\begin{align}\label{eq:H alpha less 1/2}
		H=&\sqrt{\frac{\rho_0}{3}}\left[B+\frac{\sqrt{3}(1-2\alpha)A}{2\rho_0^{\frac12-\alpha}}(t-t_\star)\right]^{\frac{1}{1-2\alpha}},\\
		\dot{H}=&\frac{A}{2}\rho_0^\alpha\left[B+\frac{\sqrt{3}(1-2\alpha)A}{2\rho_0^{\frac12-\alpha}}(t-t_\star)\right]^{\frac{2\alpha}{1-2\alpha}}.\label{eq:dotHLRLSBR}
	\end{align}
	For $0<\alpha<\frac12$, both quantities tend to infinity at the infinite asymptotic future, thus, leading to the occurrence of a LR abrupt event. On the contrary, $\dot{H}$ remains constant for $\alpha=0$, or shrinks to zero when $\alpha<0$. This behavior corresponds to a final fate \textit{\`a la} LSBR, see reference \cite{LSBR}. Note that since the DE density grows unbounded in both abrupt events, the LR and the LSBR, see equation (\ref{eq:rho}), this distinction could not be done analyzing only the asymptotic behavior of $a$ and $\rho$. Additionally note that given some entire number $n$, all the higher order derivatives of $H$ up to the $n$-th order diverge when $\alpha=(n-1)/2n$.
	
	Therefore, we conclude that the phantom DE model described by the EoS  (\ref{eq:EoS}) with a positive parameter $A$ entails a great variety of cosmological singularities and abrupt events. We summarized the results from reference \cite{EOSalpha2table} and our new findings here in Table \ref{tab:EoSalpha}. We will present a more detailed analysis in another work.

	\begin{specialtable}[H]
		\caption{Classification of the singularities and abrupt events found for the phantom DE model given by the general EoS (\ref{eq:EoS}), see reference \cite{EOSalpha2table}, accordingly to the value of the parameter $\alpha$, the time of occurrence of the singular behavior $t_s$, the Hubble rate $H$ and its cosmic time derivative $\dot{H}$. Please note that since the BR and grand rip \cite{Fernandez-Jambrina:2014sga} singularities have qualitatively the same behavior in terms of those quantities, we do not address the possible differences between both events in the following classification. Hence, we keep the term ``big rip'' for both of them.
			\label{tab:EoSalpha}}
		\setlength{\tabcolsep}{6.4mm}
		\begin{tabular}{cccccl}
			\toprule
			\boldmath{$\alpha$}	& \boldmath{$t_s$}	& \boldmath{$a$} & \boldmath{$H$} & \boldmath{$\dot{H}$} & \textbf{Event}\\
			\midrule
			$1<\alpha$	 & Finite   & Finite   & $\infty$ & $\infty$ & Big freeze\\
			$\frac12<\alpha\leq1$& Finite   & $\infty$ & $\infty$ & $\infty$ & Big rip\\
			$0<\alpha\leq\frac12$& $\infty$ & $\infty$ & $\infty$ & $\infty$ & Little rip\\
			$\alpha\leq0$& $\infty$ & $\infty$ & $\infty$ & Finite   & LSBR\\
			\bottomrule
		\end{tabular}
	\end{specialtable}
	
	
	\subsection{Cosmological Constraints\label{sec:BR_LR_LSBR}}
	
	Since this review is mainly devoted to the analysis of the classical and quantum fate of the BR, the LR and the LSBR events in metric $f(R)$  theories of gravity, we should now restrict our attention to some particular phantom DE models on which to apply the reconstruction techniques in the next sections. Therefore, hereafter we shall consider only the following phantom DE models when addressing the occurrence of these cosmic events:
	\begin{itemize}
		\item For the BR singularity we consider the phantom DE model with a constant EoS parameter $w<-1$ \cite{BR}. This model corresponds to the choice of $\alpha=1$ in the more general EoS (\ref{eq:EoS}) studied in the previous section.
		\item For a DE model with a future LR abrupt event, we select the EoS for DE described in reference \cite{LR}, which corresponds to the case $\alpha=\frac12$ in (\ref{eq:EoS}).
		\item For a universe doomed to evolve towards a LSBR abrupt event, we consider the DE content to be described by the EoS (\ref{eq:EoS}) with $\alpha=0$. This LSBR was first introduced in reference \cite{LSBR}.
	\end{itemize}
	\textls[+25]{Note that the BR model here considered is just a subcase of the more general $w$CDM scenario, which has been thoroughly analyzed in the literature (see, for example,\linebreak \mbox{references \cite{Planck1,Planck2,Abbott:2018wog, Lopez-Corredoira:2016pwg, wCDM1,wDCM2,Addison:2013haa,Santos:2015nua,Bonilla:2017ygx}}, among others). More importantly, these specific phantom DE models have been shown to be compatible with the current observational data; }see, for instance, references  \cite{Planck1,Planck2,Abbott:2018wog,CosmoConstraints2, CosmoConstraints, LR, Bouali:2021upl}. Therefore, our own universe may evolve towards some of these singular fates. 
	
	In the following, for the sake of concreteness, we shall consider the cosmological constraints obtained in reference \cite{CosmoConstraints} when binding those DE models with observational data. Thus, the results presented in the incoming sections are subjected to the observational constraints on the parameter $A$ presented in reference \cite{CosmoConstraints}. These constraints are summarized in Table \ref{tab:Cosmoconstraints}. Please note that the small values for $A$ there obtained suggest that tiny deviations from the $\Lambda$CDM scenario are, in fact, the observationally preferred situation today \cite{CosmoConstraints}. Nevertheless, we recall that the asymptotic evolution of these DE models is not that of a de Sitter universe, since the corresponding DE pressure and energy density do not converge to a constant value but diverge. In the next section, we will obtain the group of metric $f(R)$ theories of gravity able to reproduce the same asymptotic expansion history, which in GR corresponds to these particular phantom DE models.
	\begin{specialtable}[H]
		\caption{Best fit found in reference \cite{CosmoConstraints} for the phantom DE models discussed in Section \ref{sec:BR_LR_LSBR}, where $A$ is dimensionless for the case of the BR, and has units of inverse of meter and inverse of square meter for the LR and LSBR models, respectively.
			\label{tab:Cosmoconstraints}}
		\setlength{\tabcolsep}{15.8mm}
		\begin{tabular}{l c c}
			\toprule
			& \boldmath{$\alpha$} & \boldmath{$A$}\\
			\midrule
			BR	 & $1$       &  $0.0276\pm0.0240$\\
			LR   & $\frac12$ &  $\left(2.75\pm1.30\right)\times10^{-28}$\\
			LSBR & $0$       &  $\left(2.83\pm4.17\right)\times10^{-54}$   \\
			\bottomrule
		\end{tabular}
	\end{specialtable}


	\section{Phantom Dark-Energy Models in \boldmath{$f(R)$} Cosmology \label{sec:fR}}
	
	For a given cosmological background evolution it is possible to find a family of alternative theories of gravity that leads to the same expansion history. The group of techniques used to perform such a background ``reconstruction'' task are dubbed as ``reconstruction methods'' (for a review of the topic see, for instance,  reference \cite{ReconNojiri99pag} and references therein).  In this part of the review, we shall focus our attention on reconstruction methods within the scheme of metric $f(R)$ theories of gravity. Hence, we look for a metric $f(R)$ theory of gravity able to reproduce the same superaccelerated expansion profile to that of the general relativistic model filled with phantom DE with the EoS (\ref{eq:EoS}), for the values of $\alpha$ selected in Section \ref{sec:BR_LR_LSBR}. It is worth noting that whereas in GR the accelerated phase is attributed to the existence of an exotic form of energy with a negative pressure (DE), in the setup of $f(R)$ theories of gravity the same expansion has a rather geometrical origin. Furthermore, as the general relativistic model with the EoS (\ref{eq:EoS})  expands the cosmos towards some future doomsdays, then the resulting metric $f(R)$ theory of gravity will lead to the same classical fate.
	For previous works on reconstruction techniques in $f(R)$ gravity see, for instance, references \cite{ReconNojiri99pag,ReconCapozziello,ReconfRgravity,ReconMGravity,ReconNojiri,ReconDunsby,ReconCarloni}. See also references \cite{LRinfHG,ReconRef1,BRfR,LSBRfR,LRfR} for successful reconstruction of phantom DE-driven riplike events in metric $f(R)$ theories of gravity.
	
	Henceforth, we shall refer to two cosmological evolutions as equivalent at the background level if the corresponding geometrical variables $H$, $\dot{H}$, $R$ and $\dot{R}$ are identical \cite{ReconRef1}. 
	In GR, the expansion of a  homogeneous and isotropic universe is ruled by the Friedmann and Raychaudhuri equations (\ref{eq:FE}) and (\ref{eq:FE2}), respectively. Accordingly, the scalar curvature reads
	\begin{equation}\label{eq:R}
		R=6\left(\dot{H}+2H^2+\frac{k}{a^2}\right)=\rho-3p,
	\end{equation}
	where $\rho$ and $p$ denotes the total energy density and pressure, respectively, these are $\rho=\rho_{DE}+\rho_M$ and $p=p_{DE}+p_{M}$. Later we will assume that they just correspond to the DE component when the cosmic matter is diluted.
	Additionally, from the continuity equation for the perfect fluid (\ref{eq:conservationRho}) 
	and the Friedmann equation (\ref{eq:FE}), it follows
	\begin{align}\label{dot rho and p}
		\dot{\rho}&=-3(p+\rho)\left(\frac{1}{3}\rho-\frac{k}{a^2}\right)^{\frac{1}{2}}   , \\
		\dot{p}&=-3(p+\rho)\left(\frac{1}{3}\rho-\frac{k}{a^2}\right)^{\frac{1}{2}}\frac{dp}{d\rho}   ,
	\end{align}
	where we have assumed $p=p(\rho)$. Consequently, the cosmic time derivative of the scalar curvature $R$ reads
	\begin{equation}\label{dot R}
		\dot{R}=-3\,(p+\rho)\left(\frac{1}{3}\rho-\frac{k}{a^2}\right)^{\frac{1}{2}}\left(1-3\frac{dp}{d\rho}\right)  .
	\end{equation}
	
	On the other hand, for the gravitational interaction being that provided by metric $f(R)$ theories of gravity, the evolution of the universe is described by the action
	\begin{eqnarray}
		S=\frac{1}{2}\int d^4x\sqrt{-g}f(R)+S_m,
	\end{eqnarray}
	where $S_m$ stands for the minimally coupled matter fields.
	In this framework, the field equations no longer coincide with (\ref{eq:FE}) and (\ref{eq:FE2}). In fact, the modified Friedmann equation~reads
	\begin{equation}\label{eq:M.F.E.}
		3H^2 \frac{df}{dR}=\frac{1}{2}\left(R\frac{df}{dR}-f\right)-3H\dot{R}\frac{d^2f}{dR^2	}-3\frac{k}{a^2}+\rho_m ,
	\end{equation}
	being $\rho_m$ the energy density of the minimally coupled matter fields. 
	Since we are interested in a metric $f(R)$ theory able to reproduce the same background cosmological expansion as a certain general relativistic model, then the preceding expression can be considered to be a differential equation for some, \textit{a priori} unknown, function $f(R)$, where the coefficients are already fixed, i.e., when the geometrical quantities involved in equation (\ref{eq:M.F.E.}) are set to be equal to those of the GR model that we want to reproduce. Thus, the background cosmological expansion of the resulting metric $f(R)$ theory of gravity will be equivalent to that provided by the general relativistic model. 
	Furthermore, as we are interested on the asymptotic future behavior of the universe, we can neglect the matter and spatial curvature contribution in equations (\ref{eq:R}) and (\ref{eq:M.F.E.}), which will be (more) quickly redshifted with the superaccelerated expansion. Therefore, we drop again the subindex DE for the energy density and pressure.
	Hence, in the following sections we present the most general solution to this reconstruction procedure when considering the EoS (\ref{eq:EoS}) for the different values of $\alpha$ discussed in Section \ref{sec:BR_LR_LSBR}. Nevertheless, the obtained $f(R)$ theories must be understood as useful asymptotic models for the study of the behavior of the universe near classical singularities, since a more realistic background reconstruction would imply the non-cancellation of the matter and spatial curvature contribution in \mbox{equations (\ref{eq:R}) and (\ref{eq:M.F.E.})}.
	
	
	\subsection{BR Singularity in $f(R)$ Gravity \label{sec:BRinfR}}
	For the BR singularity we considered the case of $\alpha=1$ in the DE general EoS (\ref{eq:EoS}). This~is
	\begin{eqnarray}
		p=-(1+A)\rho.
	\end{eqnarray}
	For that specific value of $\alpha$, the cosmic time derivative of the Hubble rate can be re-expressed~as
	\begin{eqnarray}
		\dot{H}=\frac32 AH^2,
	\end{eqnarray}
	compare equations (\ref{eq:H alpha 1}) and (\ref{eq:dotH alpha 1}).
	Consequently, the scalar curvature and its cosmic time derivative reduces to
	\begin{align}
		R=&3(4+3A)H^2,\\
		\dot{R}=&9(4+3A)A H^3.
	\end{align}
	Therefore, the modified Friedmann equation (\ref{eq:M.F.E.}) reads
	\begin{eqnarray}
		R^2 f_{RR}-\frac{2+3A}{6A}Rf_R+\frac{4+3A}{6A}f=0,
	\end{eqnarray}
	where we have used the notation $f_R\coloneqq df/dR$ and $f_{RR}\coloneqq d^2f/dR^2$.
	The most general solution for the function $f(R)$ was already obtained in reference \cite{ReconRef1}. That is
	\begin{align}\label{eq:BRfR}
		f(R)=c_+R^{\gamma_+} + c_-R^{\gamma_-},
	\end{align}
	being $c_+$ and $c_-$ arbitrary integration constants and
	\begin{align}\label{eq:BRgamma}
		\gamma_\pm\coloneqq\frac12\left\{1+\frac{2+3A}{6A}\pm\sqrt{\left[1+\frac{2+3A}{6A}\right]^2-\frac{2(4+3A)}{3A}}\right\}.
	\end{align}
	In general, the parameter $\gamma_\pm$ may take complex values. However, for $\big\lbrace A\leq(10-4\sqrt{6})/3\big\rbrace\cup\big\lbrace A\geq(10+4\sqrt{6})/3\big\rbrace$, both branches are real valued. Please note that this is precisely the case for the values of $A$ showed in Table \ref{tab:Cosmoconstraints}. Hence, when the observational \mbox{constraints \cite{CosmoConstraints}} on the model are taken into account, both branches of the parameter $\gamma_\pm$ are real valued.

	\subsection{The LR in $f(R)$ Gravity\label{sec:LRinfR}}
	For the LR abrupt event, we consider here the EoS (\ref{eq:EoS}) for DE with $\alpha=\frac12$ in (\ref{eq:EoS}). This is
	\begin{eqnarray}\label{eq:EoSLR}
		p=-\rho-A\sqrt{\rho}.
	\end{eqnarray}
	Note that for this choice of $\alpha$, the Hubble rate in equation (\ref{eq:HLR}) is an exponential function on the cosmic time. Subsequently, its cosmic time derivative is proportional to itself. Thus, for the sake of simplicity, we denoted by $\beta$ that proportionality constant. That is
	\begin{equation}
		\dot{H}=\beta H,
	\end{equation}
	where we have defined
	\begin{eqnarray}
		\beta\coloneqq\frac{\sqrt{3}}{2}A.
	\end{eqnarray}
	The curvature scalar and its cosmic time derivative in terms of the Hubble rate are
	\begin{align}\label{eq:R(H)LR}
		R&=6 H\left(\beta+2H\right),\\
		\dot{R}&=6\beta H\left(\beta +4H\right).\label{eq:dotR(H)LR}
	\end{align}
	As noted in reference \cite{LRinfHG} (se also \cite{LRfR}), the modified Friedmann equation (\ref{eq:M.F.E.}) simplifies when rewritten in terms of $H$. Thus, we need to solve
	\begin{align}
		\beta H^2\left(\beta+4H\right)f_{HH}-\left[4\beta H^2+H\left(\beta+H\right)\left(\beta+4H\right)\right]f_H+\left(\beta+4H\right)^2f=0,
	\end{align}
	where $f_H\coloneqq df/dH$, $f_{HH}\coloneqq d^2f/dH^2$, and $\rho_m$ and $k$ has been neglected, as discussed before. The most general solution to the above differential equation is \cite{LRinfHG}
	\begin{align}\label{sol:LRinf(R)}
		f(H)=&C_1\left(H^4-5\beta H^3+2\beta^2H^2+2\beta^3H\right)+C_2\bigg[\beta H\left(\beta^2+4\beta H-H^2\right)e^{\frac{H}{\beta}}\nonumber\\
		& +\left(H^4-5\beta H^3+2\beta^2H^2+2\beta^3H\right)\textup{Ei}\left(\frac{H}{\beta}\right)\bigg],
	\end{align}
	being $C_1$ and $C_2$ integration constants and Ei the exponential integral function (see definition, e.g., in 5.1.2 of reference \cite{libroFunciones}).
	To obtain the final $f(R)$ expression, the relation (\ref{eq:R(H)LR}) must be inverted. This is
	\begin{equation}\label{eq:H(R)}
		H=\frac{1}{12}\left(-3\beta\pm\sqrt{9\beta^2+12R}\right).
	\end{equation}
	At a first sight, it seems that this transformation may be non-univocally defined. However, since the phantom DE density (\ref{eq:rho}) increases with the expansion of the universe towards the LR abrupt event, whereas the matter content is quickly diluted, then the corresponding Hubble rate (\ref{eq:HLR}) and scalar curvature (\ref{eq:R(H)LR}) also grow. Hence, only the positive branch in the preceding expression applies. Therefore,
	\begin{align}\label{eq:LRfR}
		f(R)=&c_1\left[27\beta^4+150\beta^2R-\beta\left(9\beta^2+12R\right)^{\frac32}+2R^2\right]\nonumber\\
		&+c_2\Bigg\{\beta\left(-3\beta^2-2R+9\beta\sqrt{9\beta^2+12R}\right)\nonumber\\
		&\times\left(-3\beta+\sqrt{9\beta+12R}\right)\exp\left(-\frac14+\frac14\sqrt{1+\frac{4R}{3\beta}}\right)\nonumber\\
		&+ \left[27\beta^4+150\beta^2R-\beta\left(9\beta^2+12R\right)^{\frac32}+2R^2\right]\textup{Ei}\left(-\frac14+\frac14\sqrt{1+\frac{4R}{3\beta}}\right)\Bigg\}
	\end{align}
	represents the most general family of metric $f(R)$ theories of gravity predicting the occurrence of that specific LR abrupt event \cite{LRinfHG} (see also \cite{LRfR}). We recall that  $\beta$, which depends on the parameter $A$, is observationally constrained in Table \ref{tab:Cosmoconstraints}.
	
	
	\subsection{The LSBR in $f(R)$ Gravity\label{sec:LSBRinfR}}
	
	The selected  DE model in this case corresponds to $\alpha=0$  in the EoS (\ref{eq:EoS}). That is
	\begin{eqnarray}
		p=-\rho-A,
	\end{eqnarray}
	where $A$ takes the value showed in Table \ref{tab:Cosmoconstraints}. In this scenario, see equation (\ref{eq:dotHLRLSBR}), the cosmic time derivative of the Hubble rate remains constant. Its value is given by
	\begin{eqnarray}
		\dot{H}=\frac{A}{2}.
	\end{eqnarray}
	Hence, the scalar curvature and its cosmic time derivative read
	\begin{align}
		R&=12H^2+3A,\label{eq:R(H) LSBR}\\
		\dot{R}&=12AH.
	\end{align}
	Consequently, the modified Friedmann equation (\ref{eq:M.F.E.}) becomes
	\begin{eqnarray}
		3A(R-3A)f_{RR}-\frac{1}{4}(R+3A)f_R+\frac12 f=0,
	\end{eqnarray}
	whose most general solution reads \cite{LSBRfR}
	\begin{align}\label{eq:LSBRfR}
		f(R)& = c_1\left(9A^2-18AR+R^2\right) +c_2 \left(\frac{R-3A}{12A}\right)^{\frac{3}{2}} 
		{}_1F_1\left(-\frac{1}{2};\frac{5}{2};\frac{R-3A}{12A}\right) ,
	\end{align}
	where $c_1$ and $c_2$ are integration constants and ${}_1F_1$ is the confluent hypergeometric functions or Kummer's function, see definition in chapter 13 of reference \cite{libroFunciones}. See Table \ref{tab:Cosmoconstraints} for the possible values of $A$ we shall consider here.
	
	
	\subsection{Viability and Local System Tests}

	It is a well-known fact that local tests  pose rather tight constraints on the metric formulation of $f(R)$ theories of gravity (see, for example, references \cite{Chiba:2006jp,Cassini,Capozziello:2005bu,Capozziello:2007ms,Capozziello:2009zz,ODwyer:2013vfo,Wang:2019svo}). Thus, any candidate for a reliable alternative to GR should pass or, somehow, evade these low-curvature-regime tests (see also \cite{Sawicki:2007tf,Hu:2007nk,Roshan:2011kz,Guo:2013fda,Naik:2018mtx,Negrelli:2020yps} for an interesting discussion). However, the metric $f(R)$ models presented here were expressly built to reproduce a high curvature regime very different from that of (an effective) $\Lambda$CDM. Please note that the contributions of  the matter (dark and baryonic) and the spatial curvature $k$  to equations (\ref{eq:R}) and (\ref{eq:M.F.E.}) have been de facto ignored. Therefore, the resulting $f(R)$ theories obtained here must be seen as useful asymptotic models for the theoretical evaluation of the quantum fate of classical singularities rather than complete proposals for viable alternatives to GR at all scales.

	
	\section{\boldmath{$f(R)$} Quantum Geometrodynamics \label{sec:QC}}
	
	Even though there is a lack of consensus on what is the correct quantum theory of gravity, the application of ordinary quantum mechanics to the universe as a whole leads to an interesting framework, known as quantum cosmology (for a review of the topic see, e.g., references \cite{Klauslibro,KlausBarbara}). Currently there are multiple proposals to quantize the cosmological background. A non-exhaustive listing of different approaches to quantum cosmology is string theory, loop quantum cosmology \cite{Bojowald:2006da,Ashtekar:2011ni}, causal dynamical \mbox{triangulation \cite{CDT1,CDT2,CDT3}} and quantum geometrodynamics \cite{DeWitt}, among other examples (see also reference \cite{Klauslibro}).
	Nonetheless, in this review we shall focus only on the latter approach, which corresponds, in fact, to one of the first attempts to quantize cosmological backgrounds \cite{DeWitt}. In the DeWitt's pioneering work \cite{DeWitt}, a quantization procedure for a closed Friedmann universe was provided, leading to the first minisuperspace model in quantum cosmology. This quantum cosmology is based on a canonical quantization with the Wheeler-DeWitt equation for the wave function $\Psi$ of the universe playing a central role \cite{DeWitt,Kuchar:1989tj,Wheeler}. The expression minisuperspace stands for a cosmological model truncated to a finite number of degrees of freedom. This  nomenclature is derived from the usage of ``superspace''  to denote the full infinite-dimensional configuration space of GR and the prefix ``mini'' for the truncated versions. Moreover, DeWitt also proposed a criterion for the avoidance of classical singularities within the quantum regime, namely the DW criterion. This is, the classical singularity is potentially avoided if the wave function of the universe vanishes in the nearby configuration space. This criterion is based on a generalization of the interpretation of the wave function in ordinary quantum theory, where the wave function is the fundamental building block for any observable. Consequently, regions of the configuration space that lay outside of the support of $\Psi$ are, therefore, irrelevant in practice
	\footnote{Alternatively to this interpretation, the wave function $\Psi$ can be linked in a heuristic way with the probability distribution. In that sense, having a vanishing wave function could be interpreted as having zero probability of reaching that point in the configuration space. Nevertheless, this interpretation is based on the existence of squared integral functions and a consistent probability interpretation of the wave function. The problem is that these assumptions would require a minisuperspace with a proper Hilbert space nature, and that is not obvious to be always doable for a quantum cosmology based on the WDW equation. 
	}. It should be noted, however, that the non-vanishing of the wave function does not necessary entail a singularity. Therefore, the DW criterion can only  be a sufficient but not necessary criterion for the avoidance of singularities. This criterion has been successfully applied in several cosmological scenarios, see,  e.g., references \cite{SingularityClasification,BouhmadiLopez:2009pu,Dabrowski:2006dd,GRLR,Albarran:2015cda,BRfR,LSBRfR,Bouhmadi-Lopez:2016dcf,Bouhmadi-Lopez:2018tel,Albarran:2018mpg} among others.
	
	
	\subsection{Modified Wheeler-DeWitt\label{sec:mWDW}}
	
	Following the ideas presented in reference \cite{Vilenkin}, the Wheeler-DeWitt equation must be adapted to the framework of $f(R)$ theories of gravity. It is worthy to note that when investigating the canonical quantization of an $f(R)$ theory we are interpreting that theory as a fundamental theory of gravity, rather than an effective framework coming from a quantum gravity proposal. As this classical theory of gravity implies the occurrence of singularities, one concludes that it must be quantized. In this section, we shall show how the formalism of quantum geometrodynamics can be adopted to perform such quantization.  The resulting scheme is known as $f(R)$ quantum geometrodynamics.

	For the gravitational interaction being that provided by metric $f(R)$ theories of gravity, cosmological models can be described by the action
	\begin{equation}
		S=\frac{1}{2}\int d^4x \sqrt{-g}f(R),
	\end{equation}
	in the so-called Jordan frame.
	For a FLRW background metric, the above action reduces to 
	\begin{equation}\label{eq:actionf(R)mWDW}
		S=\frac{1}{2}\int dt \ \mathcal{L}(a,\dot{a},\ddot{a}),
	\end{equation} 
	where the point-like Lagrangian reads
	\begin{equation}
		\mathcal{L}(a,\dot{a},\ddot{a})=\mathcal{V}_{(3)}\,a^3f(R),
	\end{equation} 
	denoting by $\mathcal{V}_{(3)}$ the spatial three-dimensional volume. 
	Please note that metric $f(R)$ theories of gravity have an additional degree of freedom when compared with GR (see Einstein's and Jordan's frame formulation in \cite{EvsJframe1,EvsJrframe2,EvsJrframe3,EvsJrframe4} and references therein).  Consequently, a new variable can be introduced for the canonical quantization of these alternative theories of gravity. Furthermore, this new variable can be selected in such a way to remove the dependence of the action (\ref{eq:actionf(R)mWDW}) on the second derivatives of the scale factor. Hence, following the line of reasoning presented in reference \cite{Vilenkin}, we select the scalar curvature, $R$, to be the new variable. Then, the action (\ref{eq:actionf(R)mWDW}) becomes
	\begin{eqnarray}
		S=\frac{1}{2}\int dt \ \mathcal{L}(a,\dot{a}, R, \dot{R}).
	\end{eqnarray}
	However, since the scalar curvature and the scale factor are not independent
	(at the classical level), their relation needs to be properly introduced in the theory via a Lagrange multiplier, $\mu$, for the classical constraint $R=R(a,\dot{a},\ddot{a})$ given in equation (\ref{eq:R}). Thence,
	\begin{equation}\label{eq:Lagrangian2}
		\mathcal{L}=\mathcal{V}_{(3)} a^3\left\lbrace f(R)-\mu\left[ R-6\left(\frac{\ddot{a}}{a}+\frac{\dot{a}^2}{a^2}+\frac{k}{a^2}\right) \right]\right\rbrace   .
	\end{equation}
	The Lagrange multiplier can be determined by varying the action with respect to the scalar curvature. This leads to
	\begin{eqnarray}
		\mu=f_R(R).
	\end{eqnarray}
	Accordingly, the point-like Lagrangian (\ref{eq:Lagrangian2}) can be reformulated as
	\begin{align}\label{eq:LmWDW}
		\mathcal{L}(a,\dot{a},R,\dot{R})=\mathcal{V}_{(3)}\bigg\{ a^3\Big[f(R)-R f_R(R)\Big]-6a^2f_{RR}(R)\dot{a}\dot{R}+6af_R(R)(k-\dot{a}^2)\bigg\}.
	\end{align}
	%
	For the sake of the quantization procedure, the derivative part of the above point-like Lagrangian can be diagonalized by the introduction of a new set of variables \cite{Vilenkin}. {These~are}
	\begin{subequations}\label{eq:Vilenkin}
		\begin{align}
			\label{eq:Vilenkin q} 
			q&\coloneqq a\sqrt{R_\star}\left(\frac{f_R}{f_{R_\star}}\right)^{\frac{1}{2}},\\
			x&\coloneqq\frac{1}{2}\ln\left(\frac{f_R}{f_{R_\star}}\right),\label{eq:Vilenkin x} 
		\end{align}
	\end{subequations}
	being $R_\star$ a constant needed for the change of variables to be well-defined. There are different proposals among the existing literature for the value of this constant, see, for instance, references \cite{Vilenkin,BRfR,LSBRfR,LRfR}. In this work we adopt the convention discussed in \mbox{reference \cite{LRfR}}, where $R_\star$ is defined as the value of the curvature scalar evaluated at some future scale factor $a=a_\star$ on which the description of the universe by means of DE only becomes appropriate. That is
	\begin{eqnarray}\label{eq:Rstar}
		R_\star\coloneqq R(a,\dot{a},\ddot{a})\Big|_{a=a_\star},
	\end{eqnarray} 
	where $R(a,\dot{a},\ddot{a})$ obeys the classical relation (\ref{eq:R}). Furthermore, for the sake of concreteness, we can safely assume $a_\star=100a_0$ hereafter. Since at that moment in the expansion the matter content will be  diluted by a factor of $10^{-6}$ with respect to the present concentration and, therefore, $\Omega_{DE}\approx1$. 
	Please note that from this definition it follows $R>R_\star$, since the scalar curvature asymptotically approaches an increasing function with the expansion of the universe. For different definitions of $R_\star$ see references \cite{Vilenkin,BRfR,LSBRfR}.
	In these new variables, equation (\ref{eq:LmWDW}) transforms into
	\begin{align}\label{eq:L(x,q)}
		\mathcal{L}(q,\dot{q}, x, \dot{x})=\mathcal{V}_{(3)}\left(\frac{R_\star f_R}{f_{R_\star}}\right)^{-\frac{3}{2}}q^3 \left[f-6f_R\frac{\dot{q}^2}{q^2} -Rf_R +6f_R\dot{x}^2+6k\frac{R_\star}{f_{R_\star}}\frac{f^2_R}{q^2}\right],
	\end{align}
	where $f$ and $f_R$ are now understood as functions of $x$. This form of the Lagrangian is already suitable for the quantization procedure.
	
	Since the kinetic part of the point-like Lagrangian has been diagonalized, the derivation of the corresponding Hamiltonian is straightforward. The conjugate momenta are
	\begin{align}
		P_q&=\frac{\partial\mathcal{L}}{\partial \dot{q}}=-12\mathcal{V}_{(3)} R_\star^{-\frac{3}{2}}f^{\frac{3}{2}}_{R_\star}f^{-\frac{1}{2}}_R q \dot{q} ,\\
		P_x&=\frac{\partial\mathcal{L}}{\partial \dot{x}}=12\mathcal{V}_{(3)} R_\star^{-\frac{3}{2}}f^{\frac{3}{2}}_{R_\star}f^{-\frac{1}{2}}_R q^3 \dot{x}.
	\end{align}
	Then, the corresponding Hamiltonian reads
	\begin{align}\label{eq:Hamiltonianqx}
		\mathcal{H}=-\mathcal{V}_{(3)}q^3\left(\frac{R_\star f_R}{f_{R_\star}}\right)^{-3/2} \left\lbrace f+6k\frac{R_\star}{f_{R_\star}}\frac{f_R^2}{q^2} -Rf_R+\frac{6R_\star^3}{(12)^2\mathcal{V}_{(3)}^2f_{R_\star}^3}\frac{f_R^2}{q^4}\left[P_q^2-\frac{P_x^2}{q^2}\right]\right\rbrace.
	\end{align}
	For the quantization of the theory, we assume the procedure
	\begin{align}
		P_q\to&-i\hbar\diffp{}{q}, \\
		P_x\to& -i\hbar\diffp{}{x}.
	\end{align} 
	Therefore, the Hamiltonian (\ref{eq:Hamiltonianqx}) of the classical system  is promoted to a quantum operator. Consequently, the classical Hamiltonian constraint becomes the modified Wheeler-DeWitt (mWDW) equation for the wave function $\Psi$ of the universe \cite{DeWitt,Vilenkin,Klauslibro}. That is
	\begin{equation}
		\mathcal{\hat{H}}\Psi=0.
	\end{equation}
	After simple manipulations, the preceding expression can be cast in the form of the hyperbolic differential equation \cite{Vilenkin}
	\begin{equation}\label{eq:MWDW}
		\left[\hbar^2q^2\diffp[2]{}{q}-\hbar^2\diffp[2]{}{x}-V(q,x)\right]\Psi(q,x)=0 ,
	\end{equation}
	where the factor-ordering parameter has been set equal to zero. The effective potential entering the preceding equation is given by
	\begin{equation}\label{eq:V}
		V(q,x)=\frac{q^4}{\lambda^2}\left(k+\frac{q^2}{6R_\star f_{R_\star}}(f-Rf_R)e^{-4x}\right)  ,
	\end{equation}
	being $\lambda\coloneqq  R_\star/(12\mathcal{V}_{(3)}f_{R_\star})$. Please note that for a given $f(R)$ expression, the variables $q$ and $x$  are univocally fixed. Then, the relation in (\ref{eq:Vilenkin x}) must be reversed to express $f$ and $Rf_R$ in terms of $x$. However, this may not always be possible analytically, limiting the non-numerical evaluation of the mWDW equation (\ref{eq:MWDW}) to some suitable group of $f(R)$ theories.  
	In the next sections, we will review the results for the wave function $\Psi$ presented in the literature when considering the $f(R)$ expressions previously found by means of the reconstruction techniques (see Sections \ref{sec:BRinfR}--\ref{sec:LSBRinfR}). We recall that those metric $f(R)$ theories of gravity lead to the same asymptotic  background expansion as their respective general relativistic phantom DE models and, therefore, they predict  (at the classical level) singular cosmological behaviors. Accordingly, we will neglect the contribution of the spatial curvature $k$ in the effective potential (\ref{eq:V}) when studying the asymptotic fate of those models, since it will be quickly redshifted with the superacceletared expansion.
	
	Before proceeding further, we want to address some comments on the structure of the mWDW equation (\ref{eq:MWDW}). First, we want to emphasize the well-known ambiguity in the factor ordering in equation (\ref{eq:MWDW}), i.e., we could have chosen a different factor ordering when applying the quantization procedure on the Hamiltonian of the classical theory, which could have led to a different wave function of the universe. According to reference \cite{Vilenkin}, a variation of the factor ordering affects only the pre-exponential factor of the semiclassical wave function. Thus, the actual value of this parameter can be considered to be unimportant for the evaluation of the DW criterion, which is the main goal of this review. In fact, the DW criterion was found to be satisfied independently of the chosen factor ordering for some particular models related to the topic of this review, see \mbox{references \cite{Albarran:2015cda,GRLR,EiBI_Albarran:2017swy,noLSBR2}}.
	On the other hand, note that the mWDW equation (\ref{eq:MWDW}) is a globally hyperbolic differential equation, i.e., the signature of the minisuperspace DeWitt metric,
	\begin{eqnarray}
		G^{AB}=\left(\begin{array}{cc}
			q^2 & 0\\
			0 & -1
		\end{array}\right),
	\end{eqnarray}
	is $(+,-)$. This is quite different to what happens in GR when the DE content is described by a minimally coupled phantom scalar field. In that case, the DeWitt metric has a positive signature and, therefore, the WDW equation is of an elliptic type. Examples of these models can be found, for instance, in references  \cite{GRLR,BouhmadiLopez:2009pu,Dabrowski:2006dd,Albarran:2015cda}. Additionally, a change of signature in the WDW equation has also been noticed in the presence of non-minimally coupled scalar fields \cite{ellipticWDW}. 
	Moreover, the change in the signature has implications in the imposition of boundary conditions. Whereas the hyperbolic equation has a wave-like solution and, therefore,  a well-posed initial value problem, a perturbation in the initial or boundary condition for an elliptic (or parabolic) equation will spread instantly over all the point in that domain.

	
	\subsection{The BR in $f(R)$ Quantum Cosmology \label{sec:mWDWBR}}
	
	In sight of the form of expression (\ref{eq:BRgamma}), only the negative branch gives the expected limit to a de Sitter universe when the parameter $A$ shrinks to zero. (On the contrary, the exponent $\gamma_+$ diverge when $A$ vanishes) Since small deviations from the EoS of a cosmological constant are the observationally preferred situation today \cite{CosmoConstraints} (see also references \cite{Planck1,Planck2,Abbott:2018wog, Lopez-Corredoira:2016pwg, wCDM1,wDCM2,Addison:2013haa,Santos:2015nua,Bonilla:2017ygx}, among others), hereafter we consider only the negative branch of the solution presented in equation (\ref{eq:BRfR}). This is, we assume $c_+=0$ in expression (\ref{eq:BRfR}). Therefore, in this section, we quantize a subclass of the more general family of metric $f(R)$ theories of gravity predicting the classical occurrence of a BR singularity. That is
	\begin{eqnarray}\label{eq:BRfR2}
		f(R)=c_-R^{\gamma_-}.
	\end{eqnarray}
	In the following of this section, we drop the subindex ``-'' for the sake of the notation but keeping in mind that the preceding expression corresponds only to one of the two independent solutions obtained from the reconstruction procedure. Additionally, note that cosmological constraints on the model \cite{CosmoConstraints} (see also references \cite{Planck1,Planck2}) satisfy the condition  $A\leq(10-4\sqrt{6})/3\approx 0.067$. Therefore, those constraints favor a real valued exponent in equation (\ref{eq:BRfR2}) [see discussion below equation (\ref{eq:BRgamma})]. Hence, hereafter we shall consider that the aforementioned exponent is real for values of $A$ of physical interest. 
	Additionally, note that the quantum fate of the BR singularity in $f(R)$ gravity has already  been  studied in the literature for some particular values for $A$, finding that the DW criterion can be satisfied, see reference \cite{BRfR}. 
	
	For the choice of the $f(R)$ gravity in (\ref{eq:BRfR2}), the change of variables in equation (\ref{eq:Vilenkin})~reads
	\begin{subequations}
		\begin{align}
			q=&a\sqrt{R_\star}\left(\frac{R}{R_\star}\right)^{\frac{\gamma-1}{2}},\\ 
			x=&\frac12\ln\left(\frac{R}{R_\star}\right)^{\gamma-1}.
		\end{align}
	\end{subequations}
	Moreover, from the evolution equations for the Hubble rate and its cosmic time derivative, and the definition adopted in expression (\ref{eq:Rstar}), it follows that the constant $R_\star$ is
	\begin{eqnarray}
		R_\star=(4+3A)100^{3A}\rho_0.
	\end{eqnarray}
	Accordingly, the effective potential (\ref{eq:V}) entering the mWDW equation (\ref{eq:MWDW}) becomes
	\begin{equation}
		V(q,x)=-\frac{\gamma-1}{6\lambda^2\gamma}e^{-Cx}q^6,
	\end{equation}
	where, for the sake of the calculation, we have adopted the notation
	\begin{eqnarray}
		C\coloneqq2\frac{\gamma-2}{\gamma-1}.
	\end{eqnarray}
	Thus, the mWDW equation (\ref{eq:MWDW}) reads
	\begin{equation}\label{eq:mWDWBR}
		\left[\hbar^2q^2\diffp[2]{}{q}-\hbar^2\diffp[2]{}{x}+\frac{\gamma-1}{6\lambda^2\gamma}e^{-Cx}q^6\right]\Psi(q,x)=0.
	\end{equation}
	However, the dependence of the effective potential on the  minisuperspace variables can be simplified when considering the following change of variables \cite{BRfR} (see also reference \cite{GRLR})
	\begin{subequations}\label{eq:BRThetaZ}
		\begin{align}
			q&=r\left(z\right)\theta,\\
			x&=z.
		\end{align}
	\end{subequations}
	Please note that this implies \cite{GRLR}
	\begin{align}\label{eq:changVarBR}
		\diffp[2]{}{q}=&r^{-2}\diffp[2]{}{\theta}, \\ 
		\diffp[2]{}{x}=&\left(\frac{r^{\prime}}{r}\right)^2\left[\theta^2\diffp[2]{}{\theta}+\theta\diffp{}{\theta}\right]-2\frac{{r^{\prime}}}{r}\theta\diffp{}{\theta}\diffp{}{z}+\left[\left(\frac{r^{\prime}}{r}\right)^2-\frac{r^{\prime\prime}}{r}\right]\theta\diffp{}{\theta}+\diffp[2]{}{z},
	\end{align}
	Then, for the choice of $r(z)=e^{Cz/6}$, the potential term will only depend on $\theta$ \cite{BRfR}. Accordingly,  the mWDW equation (\ref{eq:mWDWBR}) simplifies to
	\begin{align}\label{eq:mWDWBR2}
		\left[\left(1-\frac{C^2}{36}\right)\hbar^2\theta^2\diffp[2]{}{\theta}-\hbar^2\diffp[2]{}{z}+\frac{C}{3}\hbar^2\theta\diffp{}{\theta}\diffp{}{z}-\frac{C^2}{36}\hbar^2\theta\diffp{}{\theta}
		+\frac{\gamma-1}{6\lambda^2\gamma}\theta^6\right]\Psi(\theta,z)=0,
	\end{align}
	where the factor $1-C^2/36$ is different from zero at least for the range of values for $A$ of our interest, i.e., $A\leq 0.067$ [see discussion below equation (\ref{eq:BRfR2})].
	Therefore, to analyze whether the BR singularity is avoided in the quantum realm by means of the DW criterion, we focus on solving the preceding equation for the wave function $\Psi$ in the configuration space near the singularity.  As we do not expect the wave function to be peaked along the classical trajectory in this regime, $R$ and $a$ may take completely independent values. Therefore, to consider a region close to the BR singularity, we should assume either $a\rightarrow\infty$
	or $R\rightarrow\infty$. Both choices imply $\theta\rightarrow\infty$, but in the former case $z$ is arbitrary whereas in the latter one $z\rightarrow\infty$.
	Please note that the divergence of the scalar curvature can be argued to be the dominant condition, from a geometric point of view, for the occurrence of the BR singularity. Thence, we shall consider both $\theta$ and $z$ going to infinity as the main parametrization of the BR singularity in the configuration space.
	Nevertheless, the results and conclusions presented in this section are independent of this choice and still hold for $\theta\to\infty$ and $z$~arbitrary.
	
	Additionally, further simplifications can be made when solving (\ref{eq:mWDWBR2}) close to the BR. By considering the third term containing the cross partial derivatives to be unimportant when $\theta\to\infty$,  the above equation can be solved via the separation ansatz 
	\begin{eqnarray}\label{eq:BRansatz}
		\Psi(\theta,z)=\sum_{\tilde{k}}b_{\tilde{k}}\chi_{\tilde{k}}(\theta)\varphi_{\tilde{k}}(z),
	\end{eqnarray}
	where $b_{\tilde{k}}$ gives the amplitude of each solution and $\tilde{k}$ is related to the associated energy. Please do not confuse $\tilde{k}$ with the spatial curvature $k$, which has been set to zero. (The validity of this approximation is analyzed in Appendix \ref{ap:BR}.)
	Under these approximations, equation (\ref{eq:mWDWBR2}) reduces to the following system of equations
	\footnote{For a different approach to the asymptotic form of the wave function $\Psi$ see reference \cite{BRfR}.}
	\begin{align}
		\hbar^2\diffp[2]{\varphi_{\tilde{k}}}{z}-\tilde{k}^2\varphi_{\tilde{k}}&=0,\\
		\left(1-\frac{C^2}{36}\right)\hbar^2\theta^2\diffp[2]{\chi_{\tilde{k}}}{\theta}-\frac{C^2}{36}\hbar^2\theta\diffp{\chi_{\tilde{k}}}{\theta}+\left(\frac{\gamma-1}{6\lambda^2\gamma}\theta^6-\tilde{k}^2\right)\chi_{\tilde{k}}&=0.\label{eq:BRchi}
	\end{align}
	The former equation can be straightforwardly worked out. The solutions are
	\begin{eqnarray}\label{eq:BRsolVarphi}
		\varphi_{\tilde{k}}(z)=d_1\exp\left(\frac{\sqrt{\tilde{k}^2}}{\hbar}z\right)+d_2\exp\left(-\frac{\sqrt{\tilde{k}^2}}{\hbar}z\right),
	\end{eqnarray}
	being $d_1$ and $d_2$ arbitrary constants. This corresponds to either exponential or trigonometric functions on $z$, depending  whether $\tilde{k}^2$ is positive or negative, respectively. On the other hand, equation (\ref{eq:BRchi}) can be solved in an exact way by means of Bessel functions, cf. 9.1.53 of reference \cite{libroFunciones}. This solution can be expressed as
	\begin{eqnarray}\label{eq:BRsolChi}
		\chi_{\tilde{k}}(\theta)=\theta^{\frac{18}{36-C^2}}\left[u_1\textup{J}_\nu\left(\frac{\tilde{\lambda}}{3\hbar}\theta^3\right)+u_2\textup{Y}_\nu\left(\frac{\tilde{\lambda}}{3\hbar}\theta^3\right)\right],
	\end{eqnarray}
	where $u_1$ and $u_2$ are integration constants, $\textup{J}_\nu$ and $\textup{Y}_\nu$ are the Bessel functions of first and second order, respectively, and
	\begin{align}
		\nu^2\coloneqq &\frac{36}{\left(36-C^2\right)^2}\left[1+4\frac{\tilde{k}^2}{\hbar^2}\left(1-\frac{C^2}{36}\right)\right],\\
		\tilde{\lambda}^2\coloneqq& \frac{6(\gamma-1)}{\lambda^2\gamma(36-C^2)}.
	\end{align}
	When near the BR singularity, i.e., for large $\theta$, the $\chi_{\tilde{k}}$ part of the wave function $\Psi$ behaves asymptotically as
	\begin{eqnarray}
		\chi_{\tilde{k}}(\theta)\approx\sqrt{\frac{6\hbar}{\pi\tilde{\lambda}}} \theta^{-\frac{3\left(24-C^2\right)}{2\left(36-C^2\right)}}\left[\tilde{u}_1\exp\left(i\frac{\tilde{\lambda}}{3\hbar}\theta^3\right)+\tilde{u}_2\exp\left(-i\frac{\tilde{\lambda}}{3\hbar}\theta^3\right)\right],
	\end{eqnarray}
	where $\tilde{u}_1$ and $\tilde{u}_2$ now depend on $\tilde{k}$, cf. 9.2.1-2 in reference \cite{libroFunciones}. Thence, the total wave function of the universe asymptotically reads
	\begin{align}
		\Psi(\theta,z)\approx&\sqrt{\frac{6\hbar}{\pi\tilde{\lambda}}} \theta^{-\frac{3\left(24-C^2\right)}{2\left(36-C^2\right)}}\sum_{\tilde{k}}b_{\tilde{k}}\left[\tilde{u}_1\exp\left(i\frac{\tilde{\lambda}}{3\hbar}\theta^3\right)+\tilde{u}_2\exp\left(-i\frac{\tilde{\lambda}}{3\hbar}\theta^3\right)\right]\nonumber\\
		&\times\left[d_1\exp\left(\frac{\sqrt{\tilde{k}^2}}{\hbar}z\right)+d_2\exp\left(-\frac{\sqrt{\tilde{k}^2}}{\hbar}z\right)\right].
	\end{align}
	As a result, for the condition of $d_1=0$ when $\tilde{k}^2$ positive, the wave function vanishes at the BR singularity. Therefore, for this boundary condition, the DW criterion is satisfied. This result points towards the avoidance of this cosmological singularity in the quantum real of $f(R)$ cosmology. Nevertheless, it should be noted that by setting $d_1=0$ we have dismissed a subgroup of solutions to the mWDW equation as unphysical. If future investigations show the importance of these ignored solutions, then it would be concluded that the DW criterion may not always be satisfied.
	
	
	\subsection{The LR in $f(R)$ Quantum Cosmology\label{sec:mWDWLR}}
	
	In this section, we address the quantum fate of the LR abrupt event in the framework of $f(R)$ quantum geometrodynamics. Previously, in Section \ref{sec:LRinfR}, it was shown that the alternative theory of gravity given by the function (\ref{eq:LRfR}) is the most general expression for a metric $f(R)$ theory of gravity which gives the same asymptotic expansion history as GR filled with  phantom DE governed by equation (\ref{eq:EoSLR}).
	Moreover, since the presence of the Ei function in the term multiplying $c_2$ spoils the analytical inversion of $x(R)$ in  \mbox{equation (\ref{eq:Vilenkin x})}, which is crucial for the exact derivation of the mWDW equation (\ref{eq:MWDW}), hereafter we consider $c_2=0$ \cite{LRfR}. Thus, we focus on the simple, still general, group of alternative metric $f(R)$ theories of gravity predicting the classical occurrence of a LR abrupt event given by \cite{LRfR}
	\begin{eqnarray}
		f(R)=c_1\left[27\beta^4+150\beta^2R-\beta\left(9\beta^2+12R\right)^{\frac32}+2R^2\right],
	\end{eqnarray} 
	where we recall that $\beta=\sqrt{3}A/2$ and the parameter $A$ is observationally constrained to approximately $2.75\times10^{-28}$ $\textup{m}^{-1}$, see Table \ref{tab:Cosmoconstraints}. Given the expression for $f(R)$, the change of variables (\ref{eq:Vilenkin}) reads
	\begin{subequations}
		\begin{align}
			q&=a\sqrt{\frac{2c_1R_\star}{f_{R_\star}}}\left(75\beta^2-9\beta\sqrt{9\beta^2+12R}+2R\right)^\frac12,\\
			x&=\frac12\ln\left[\frac{2c_1}{f_{R_\star}}\left(75\beta^2-9\beta\sqrt{9\beta^2+12R}+2R\right)\right],\label{eq:Vilenkin x LRinfR}
		\end{align}
	\end{subequations}
	being $f_{R_\star}=2c_1\left(75\beta^2-9\beta\sqrt{9\beta^2+12R_\star}+2R_\star\right)$. The definition of the constant $R_\star$ needed for the above change of variables to be well-defined was previously discussed for the most general case, see equation (\ref{eq:Rstar}). Consequently, from equations (\ref{eq:R(H)LR}) and (\ref{eq:HLR}) it follows \cite{LRfR}
	\begin{align}\label{eq:RstarLR}
		R_\star=4\rho_0\left(1+\beta\sqrt{\frac{3}{\rho_0}}\ln100\right)^2+
		6\beta\sqrt{\frac{\rho_0}{3}}\left(1+\beta\sqrt{\frac{3}{\rho_0}}\ln100\right).
	\end{align}
	
	On the other hand, for the purpose of computing the effective potential entering the mWDW equation, the inverse of the relation (\ref{eq:Vilenkin x LRinfR}) applies. This is 
	\begin{align}
		R(x)=84\beta^2+\frac{1}{4c_1}f_{R_\star}e^{2x}\pm54\beta\sqrt{\frac{1}{48c_1}f_{R_\star}e^{2x}+2\beta^2}.
	\end{align}
	However, only the positive branch is compatible with $R$ being an increasing function of $x$ and $R > R_\star$. Thus, we choose the positive sign in the preceding expression. Thereafter, the effective potential in the mWDW equation becomes \cite{LRfR}
	\begin{eqnarray}
		V(q,x)=-U(x)q^6 ,
	\end{eqnarray}
	where
	\begin{align}\label{eq:ULR}
		U(x)=&\tilde{U}\Bigg\lbrace1+1644\frac{c_1\beta^2}{f_{R_\star }}e^{-2x}+205992\frac{c_1^2\beta^4}{f_{R_\star }^2}e^{-4x}\nonumber \\
		&+36\beta e^{-x} \left(1+336\frac{c_1\beta^2}{f_{R_\star }}e^{-2x}\right)\sqrt{\frac{3c_1}{f_{R_\star}}\left(1+96\frac{c_1\beta^2}{f_{R_\star }}e^{-2x}\right)}\nonumber\\
		&-12\beta\sqrt{\frac{3c_1}{f_{R_\star}}} e^{-x}\left[1+330\frac{c_1\beta^2}{f_{R_\star} }e^{-2x}+18\beta e^{-x}\sqrt{\frac{3c_1}{f_{R_\star}}\left(1+96\frac{c_1\beta^2}{f_{R_\star }}e^{-2x}\right)}\right]\nonumber\\
		&\times\left[1+339\frac{c_1\beta^2}{f_{R_\star}}e^{-2x}+18\beta e^{-x}\sqrt{\frac{3c_1}{f_{R_\star}}\left(1+96\frac{c_1\beta^2}{f_{R_\star }}e^{-2x}\right)}\right]^\frac12\Bigg\rbrace,
	\end{align}
	being $\tilde{U}\coloneqq f_{R_\star}/(48c_1\lambda^2R_\star)$ a constant.
	Contrary to the previous section, a change of variables such as (\ref{eq:BRThetaZ}) will no longer be able to make the potential one-variable-dependent and, therefore, a separation ansatz such as (\ref{eq:BRansatz}) is not appropriate in this case.
	Instead, note that the $U(x)$ part of the effective potential converges very quickly to a constant value when the model is observationally constrained. This feature suggests an adiabatic semiseparability type ansatz for the wave function of the universe. This is based on the so-called Born-Oppenheimer (BO) ansatz originally formulated in the context of molecular physics \cite{BOoriginal} and first introduced in the framework of quantum cosmology in references \cite{ClausBO1,BOenWDW,ClausBO2}. Furthermore, in reference \cite{LRfR} we have argued that the scalar curvature can be considered more fundamental from a geometrical point of view than the scale factor, justifying that the following ansatz \textit{\`a la} Born-Oppenheimer should apply
	\begin{eqnarray}\label{eq:BOLR}
		\Psi(q,x)=\sum_{\tilde{k}}b_{\tilde{ k}}\chi_{\tilde{ k}}(q,x)\varphi_{\tilde{ k}}(x),
	\end{eqnarray}
	where we recall that $x$ depends only on $R$, whereas $q$ contains both $R$ and $a$, see the definitions in (\ref{eq:Vilenkin}). Additionally, $b_{\tilde{k}}$ represents the amplitude of each solution and $\tilde{k}$ is related to the associated energy. (Please do not confuse $\tilde{k}$ with the spatial curvature $k$ which has been set to zero). As a result of this ansatz, the mWDW equation (\ref{eq:MWDW}) becomes
	\begin{align}\label{eq:LRBOmWDW}
		\hbar^2q^2\varphi_{\tilde{ k}}\diffp[2]{\chi_{\tilde{ k}}}{q}-\hbar^2\varphi_{\tilde{ k}}\diffp[2]{\chi_{\tilde{ k}}}{x}-2\hbar^2\diffp{\chi_{\tilde{ k}}}{x}\diff{\varphi_{\tilde{ k}}}{x}-\hbar^2\chi_{\tilde{ k}}\diff[2]{\varphi_{\tilde{ k}}}{x}+U(x)q^6\chi_{\tilde{ k}}\varphi_{\tilde{ k}}=0.
	\end{align}
	The contribution of the second and third terms in the above expression can be neglected by virtue of the adiabatic assumption
	\footnote{The validity of this approximation is checked in Appendix \ref{ap:BO}.}. Therefore, equation (\ref{eq:LRBOmWDW}) implies
	\begin{align}
		\hbar^2\diff[2]{\varphi_{\tilde{ k}}}{x}-\tilde{ k}^2\varphi_{\tilde{ k}}&=0,\label{eq:LR BO varphi x}\\
		\hbar^2q^2\diffp[2]{\chi_{\tilde{ k}}}{q}+\left[U(x)q^6-\tilde{ k}^2\right]\chi_{\tilde{ k}}&=0.\label{eq:LR BO chi qx}
	\end{align}
	The former equation can be directly solved. The solutions are exponential and trigonometric functions on $x$, depending on the sign of $\tilde{k}^2$. These are \cite{LRfR}
	\begin{align}\label{eq:LR varphi x}
		\varphi_{\tilde{ k}}(x)=d_1\exp\left(\frac{\sqrt{\tilde{ k}^2}}{\hbar}x\right)+d_2\exp\left(-\frac{\sqrt{\tilde{ k}^2}}{\hbar}x\right), 
	\end{align}
	where $d_1$ and $d_2$ are integration constants. The solutions for $\chi_{\tilde{k}}$, on the other hand, are obtained assuming that the potential $U(x)$ behaves as a quasiconstant. Please note that this approximation is based on the fact that $U(x)$ converges very quickly to a constant value when observational constraints on the parameter $A$ are taken into account. Hence, the solutions are \cite{LRfR}
	\begin{eqnarray}\label{eq:LR chi qx}
		\chi_{\tilde{ k}}(q,x)= \sqrt{q}\left[u_1 \textup{J}_{\frac{1}{6}\sqrt{1+\frac{4\tilde{ k}^2}{\hbar^2}}}\left(\frac{\sqrt{U(x)}}{3\hbar}q^3\right)+u_2 \textup{Y}_{\frac16\sqrt{1+\frac{4\tilde{ k}^2}{\hbar^2}}}\left(\frac{\sqrt{U(x)}}{3\hbar}q^3\right)\right],
	\end{eqnarray}
	being $u_1$ and $u_2$ integration constants, cf. 9.1.53 of \cite{libroFunciones}. Thus, near the LR abrupt event, when both $q$ and $x$ diverge
	\footnote{This corresponds to $R\to\infty$ and $a$ arbitrary, which was argued to be the main condition for the appearance of a curvature singularity. Nevertheless, the results in this section would not change if we had considered both $a,R\to\infty$ instead.}, the resulting wave function of the universe becomes
	\begin{align}\label{eq:LR Psi}
		\Psi(q,x)\approx& \sqrt{\frac{6\hbar}{\pi}}\frac{1}{U(x)^{\frac14}q}\sum_{\tilde{k}}b_{\tilde{k}}\left[\tilde{u}_1\exp\left(i\frac{\sqrt{U(x)}}{3\hbar}q^3\right)+\tilde{u}_2\exp \left(-i\frac{\sqrt{U(x)}}{3\hbar}q^3\right)\right]\nonumber\\
		&\times\left[d_1\exp\left(\frac{\sqrt{\tilde{ k}^2}}{\hbar}x\right)+d_2\exp\left(-\frac{\sqrt{\tilde{ k}^2}}{\hbar}x\right)\right],
	\end{align}
	where the integration constants $\tilde{u}_1$ and $\tilde{u}_2$ now depend on $\tilde{k}^2$, see discussion in \mbox{reference \cite{LRfR}}. Since $U(x)>0$ and tends to a constant value when $x$ grows, the wave function cancels at the LR abrupt event when one of the integrations constants is set to zero. This is $d_1 = 0$ when $\tilde{k}^2$ is positive. Hence, the DeWitt criterion can be, indeed, satisfied. This points towards the avoidance of the LR abrupt event in the quantum realm of metric $f(R)$ theories of gravity. Nevertheless, as discussed for the case of the BR in the previous section, the fulfilment of the DW criterion is conditioned to the cancellation of one of the integration constants in $\Psi$. Accordingly, if future investigations claim for the physical importance of the dismissed solution, then it would have to be concluded that the DW criterion may not always be satisfied.
	
	
	\subsection{The LSBR in $f(R)$ Quantum Cosmology\label{sec:mWDWLSBR}}
	
	Finally, we present the quantum fate of the LSBR abrupt event predicted in the metric $f(R)$ theories of gravity (\ref{eq:LSBRfR}) obtained in Section \ref{sec:LSBRinfR}. Moreover, following a line of reasoning similar to that presented in the previous sections, we consider $c_2=0$ in (\ref{eq:LSBRfR}) as a necessary assumption to analytically obtain the corresponding mWDW equation.
	Therefore, we focus on the simple, still general, $f(R)$ cosmological model exhibiting a future LSBR abrupt event given by
	\begin{equation}\label{eq:LSBR mWDW fR}
		f(R)=c_1 \left(9A^2-18AR+R^2\right).
	\end{equation}
	This model has been already studied in the $f(R)$ quantum geometrodynamics approach, see reference \cite{LSBRfR}. In the next part of this section, we not only review the conclusion there presented but also provide a different, less restrictive and more general, approximation when solving for the wave function.

	For the $f(R)$ gravity model (\ref{eq:LSBR mWDW fR}), the change of variables (\ref{eq:Vilenkin}) reads
	\begin{subequations}
		\begin{align}
			q=&a\sqrt{\frac{2c_1R_\star}{f_{R_\star}}}\left(R-9A\right)^{\frac{1}{2}},\\ x=&\frac{1}{2}\ln\left[\frac{2c_1}{f_{R_\star}}\left(R-9A\right)\right], 
		\end{align}
	\end{subequations}
	being $f_{R_\star}=2c_1\left(R_\star-9A\right)$.
	Following the spirit for a physically meaningful definition of the constant $R_\star$ given in equation (\ref{eq:Rstar}), and in view of equations (\ref{eq:H alpha less 1/2}) and (\ref{eq:R(H) LSBR}), we use
	\begin{eqnarray}
		R_\star= 4\rho_0+3A\left[1+4\ln\left(100\right)\right],
	\end{eqnarray}
	where we recall that $a_\star=100a_0$ has been set, for the sake of concreteness, as the moment in the expansion history from which we can safely assume that DE is the only relevant component of the universe [see discussion below equation (\ref{eq:Rstar})].
	
	Subsequently, the mWDW equation for the  particular $f(R)$ expression considered in equation (\ref{eq:LSBR mWDW fR}) reads
	\begin{equation}\label{eq:LSBR mWDW}
		\left[\hbar^2q^2\diffp[2]{}{q}-\hbar^2\diffp[2]{}{x}+W(x)q^6\right]\Psi(q,x)=0,
	\end{equation}
	where $W(x)$ is 
	\begin{eqnarray}\label{eq:WLSBR}
		W(x)=\tilde{W}\left(1+36\frac{c_1A}{f_{R_\star}}e^{-2x}+288\frac{c_1^2A^2}{f_{R_\star}^2}e^{-4x}\right),
	\end{eqnarray}
	with $\tilde{W}\coloneqq f_{R_\star}/(24c_1\lambda^2R_\star)$ a constant.
	It should be noted that (\ref{eq:LSBR mWDW}) resembles the form of the corresponding mWDW equation for the LR case presented in the previous section. In fact, the $W(x)$ part of the effective potential also converges quickly to a constant value when $A$ is observationally constrained. Therefore, we consider here the very same BO approximation discussed in the previous section
	\footnote{Please note that in reference \cite{LSBRfR} we have used a different ansatz for solving the mWDW equation (\ref{eq:LSBR mWDW}). Since the exponential terms appearing in (\ref{eq:WLSBR}) are strongly suppressed not only by the observational value of the parameter $A$, but also by the divergence of $x$, we have considered only the asymptotic value of the function $W(x)$, i.e., we have approached the effective potential in (\ref{eq:LSBR mWDW}) to asymptotically depend only on the variable $q$. However, as argued in \cite{LRfR}, subdominant contributions to $W(x)$ have imprints in the shape of $\Psi$ near the abrupt event. Those subdominant contributions are, in fact, important when comparing different wave functions that share a common asymptotic regime.}. That is ansatz (\ref{eq:BOLR}),
	\begin{eqnarray}
		\Psi(q,x)=\sum_{\tilde{k}}b_{\tilde{ k}}\chi_{\tilde{ k}}(q,x)\varphi_{\tilde{ k}}(x),
	\end{eqnarray}
	where $b_{\tilde{k}}$ represents the amplitude of each solution and $\tilde{k}$ is related to the associated energy, do not confuse it with the spatial curvature $k$ which has been neglected.

	Accordingly to the results presented in the previous section, \textit{mutatis mutandis}, the total wave function of the universe near the LSBR abrupt event reads
	\begin{align}\label{eq:LSBR Psi}
		\Psi(q,x)\approx& \sqrt{\frac{6\hbar}{\pi}}\frac{1}{W(x)^{\frac14}q}\sum_{\tilde{k}}b_{\tilde{k}}\left[\tilde{u}_1\exp\left(i\frac{\sqrt{W(x)}}{3\hbar}q^3\right)+\tilde{u}_2\exp \left(-i\frac{\sqrt{W(x)}}{3\hbar}q^3\right)\right]\nonumber\\
		&\times\left[d_1\exp\left(\frac{\sqrt{\tilde{ k}^2}}{\hbar}x\right)+d_2\exp\left(-\frac{\sqrt{\tilde{ k}^2}}{\hbar}x\right)\right],
	\end{align}
	where $\tilde{u}_1$ and $\tilde{u}_2$ now depend on $\tilde{k}^2$. The validity of the BO approximation in this case is verified in Appendix \ref{ap:BO}. Since $W(x)>0$ and tends to a positive constant value when $x$ grows,  the wave function cancels at the LSBR abrupt event when one of the integrations constants is set to zero. This is $d_1 = 0$ when $\tilde{k}^2$ is positive. Hence, the fulfilment of the DeWitt criterion points towards the avoidance of the LSBR abrupt event in the quantum realm of metric $f(R)$ theories of gravity. Nevertheless, as stated before, if future investigations find out the importance of the dismissed solutions, then it would be concluded that the DW criterion may not always be satisfied.

	
	\section{Conclusions}
	
	The BR, LR and LSBR are cosmic curvature doomsdays predicted in some cosmological models where the superaccelerated expansion of the universe leads to the disintegration of all bounded structures and, ultimately, to the tear down of space–time itself. Within the context of GR, these cosmological catastrophes are intrinsic to phantom DE models, although a phantom fluid may also induce other cosmic singularities. For a FLRW background, these events can be characterized by the behavior of the scale factor $a$, the Hubble rate $H$ and its cosmic time derivative $\dot{H}$ near the singular fate, see Table \ref{tab1}. More importantly, some of these models have been shown to be compatible with the current observational data, see, for instance, reference \cite{CosmoConstraints}. Therefore, our own universe may evolve towards some of these (classical) doomsdays. Nevertheless, quantum gravity effects can ultimately become significant and smooth out, or even avoid, the occurrence of these classically predicted singularities. In that sense, quantum cosmology is the natural framework for addressing the quantum fate of cosmological singularities. Among the different approaches to quantum cosmology, we have focused on the canonical quantization of cosmological background due to DeWitt's pioneering paper \cite{DeWitt}. In this scheme, the DW criterion can be understood as a sufficient but not necessary condition for the avoidance of a  classical singularity. Thus, the classical singularity is potentially avoided if the wave function of the universe vanishes in the nearby configuration space. This criterion has been successfully applied in GR for the phantom DE models considered in Section \ref{sec:BR_LR_LSBR}, which predict a classical fate \textit{\`a la} BR, LR and LSBR, see references \cite{Dabrowski:2006dd,GRLR,Albarran:2015cda}. Consequently, this hints towards the avoidance of these cosmological doomsdays for those specific phantom DE~models.
	
	On the other hand, since the background late-time classical evolution can be equivalently described in the context of GR or within the framework of alternative theories of gravity, it is interesting to wonder whether the DW criterion is still fulfilled in the quantum realm of a different underlying theory of gravity. To find that out, reconstruction methods can be applied to obtain the general group of alternative theories of gravity able to reproduce the same expansion history as that of a given general relativistic model. Thus, in Section \ref{sec:fR}, we have collected the different $f(R)$ theories of gravity found among the literature that produce the same asymptotic background expansion as the phantom DE models summarized in the preceding Section \ref{sec:BR_LR_LSBR}. Consequently, these $f(R)$ models predict the classical occurrence of a BR, LR or a LSBR abrupt event. For the evaluation of the quantum fate of these classical models, we have followed the $f(R)$ quantum geometrodynamics approach introduced by Vilenkin \cite{Vilenkin}, where the Wheeler-DeWitt equation is adapted for the case of $f(R)$ theories of gravity. Within this framework, the application of the DW criterion has been discussed, showing that it is possible to find solutions to the mWDW equation with a vanishing wave function for the aforementioned cosmic events. Therefore, as it happens when the gravitational interaction is that provided by GR, this result hints towards the avoidance of these cosmological doomsdays within the scheme of metric $f(R)$ theories of gravity \cite{BRfR,LRfR,LSBRfR}. 
	
	It should be noted, however, that the validity of the wave functions obtained in this review is subject to the fulfilment of the conditions discussed in the appendices. Thus, for the avoidance of the BR singularity, for instance, we found our approximations to the wave function of the universe to be legit only for very tiny values of the parameter $A$, indeed smaller even than those found in reference \cite{CosmoConstraints} where a strong ansatz was imposed on the possible values of $w$. Of course, by relaxing such an ansatz tinier values of $A$ are compatible with cosmological observations; see, for example, references  \cite{Planck1,Planck2,Abbott:2018wog} among others.
	On the other hand, it should be also noted that we have applied the quantization procedure only to some particular solutions to the reconstruction method, i.e., we have quantized just some specific $f(R)$ functions from the more general family of metric $f(R)$ gravity found to classically predict a future fate \textit{à la} BR, LR or LSBR.
	Furthermore, certain boundary conditions have also been imposed when solving for $\Psi$ to obtain vanishing solutions at the singular points. These conditions typically involve the cancellation of one of the integration constants. Therefore, a whole subgroup of solutions to the mWDW equation has been disregarded as unphysical. If future investigations show the importance of the dismissed solutions, then it would be concluded that the DW criterion may not always be fulfilled for solutions of physical interest. 
	
	The pioneering investigations presented in this paper have raised additional questions that should be addressed. From a classical point of view, the metric $f(R)$ theories under consideration should be further analyzed, investigating their compatibility with different observational constraints, as those coming from Solar System tests. On the other hand, it would be also interesting to consider whether semiclassical effects may be important when approaching the quantum realm and, in that case, how they will affect the disintegration of bounded structures. Furthermore, we should highlight that strictly speaking, the DW criterion is not a sufficient condition to guarantee the avoidance of classical singularities. To have a complete analysis, one should obtain the  expectation values of the relevant operators. Nonetheless, the calculation of expectation values and probability  distributions is related to various open questions in quantum cosmology that still must be properly addressed in the framework of quantum geometrodynamics. Specifically, the correct  boundary or initial conditions, the Hilbert space structure and the classical-quantum correspondence, see reference \cite{Klauslibro}. Finally, we hope that the topics summarized in this work and the presented open questions will motivated further investigations.

	\vspace{6pt} 
	
	
	
	\authorcontributions{All authors contributed equally to this review. All authors have read and agreed to the published version of the manuscript.}
	
	\funding{The research of T.B.V. and P.M.-M. is supported by MINECO (Spain) Project No. PID2019-107394GB-I00 (AEI/FEDER, UE).
		T.B.V. also acknowledge financial support from Project No. FIS2016-78859-P (AEI/FEDER, UE) through Grant No. PAII46/20-08/2020-03, and from Universidad Complutense de Madrid and Banco de Santander through Grant No. CT63/19-CT64/19.
		The research of M.B.-L. is supported by the Basque Foundation of Science Ikerbasque. She also would like to
		acknowledge the partial support from the Basque government through Project No. IT956-16 (Spain) and MINECO through  Project No. FIS2017-85076-P (AEI/FEDER, UE).}

	\conflictsofinterest{The authors declare no conflict of interest.} 
	
	
	\abbreviations{The following abbreviations are used in this manuscript:\\
		
		\noindent 
		\begin{tabular}{@{}ll}
			$\Lambda$CDM& Lambda cold dark matter\\
			BF   & Big freeze\\
			BR   & Big rip\\
			DE   & Dark energy\\
			DM   & Dark matter\\
			DW   & DeWitt (criterion)\\
			EoS  & Equation of state\\
			FLRW & Friedmann-Lema\^itre-Robertson-Walker\\
			GR   & General relativity\\
			LR   & Little rip \\
			LSBR & Little sibling of the big rip \\
			WDW  & Wheeler-DeWitt (equation)\\
			mWDW & Modified Wheeler-DeWitt (equation)\\
			$w$CDM & $w$ cold dark matter\\
	\end{tabular}}
	
	\appendixtitles{yes} 
	\appendixstart
	\appendix
	\section{Validity of the wave function used to describe the BR\label{ap:BR}}
	
	When solving the mWDW equation for the BR singularity, we have neglected the contribution of the third term in (\ref{eq:mWDWBR2}). This approach is valid as long as the corresponding solutions satisfy
	\begin{eqnarray}
		\frac{C}{3}\hbar^2\theta\diffp{\chi_{\tilde{k}}}{\theta}\diffp{\varphi_{\tilde{k}}}{z}\ll\left(1-\frac{C^2}{36}\right)\hbar^2\theta^2\varphi_{\tilde{k}}\diffp[2]{\chi_{\tilde{k}}}{\theta}, \, \hbar^2\chi_{\tilde{k}}\diffp[2]{\varphi_{\tilde{k}}}{z}, \, \frac{C^2}{36}\hbar^2\theta\varphi_{\tilde{k}}\diffp{\chi_{\tilde{k}}}{\theta},\, \frac{\gamma-1}{6\lambda^2\gamma}\theta^6\chi_{\tilde{k}}\varphi_{\tilde{k}}.
	\end{eqnarray}
	According to that approximation, the solutions for $\chi_{\tilde{k}}$ and $\varphi_{\tilde{k}}$ are presented in \mbox{equations \break(\ref{eq:BRsolVarphi})} and (\ref{eq:BRsolChi}). Therefore, the terms we kept in equation (\ref{eq:mWDWBR2}) are
	\begin{small}
	\begin{align}
		\left(1-\frac{C^2}{36}\right)\hbar^2\theta^2\varphi_{\tilde{k}}\diffp[2]{\chi_{\tilde{k}}}{\theta}&\approx -\sqrt{\frac{6\tilde{\lambda}^3\hbar}{\pi}}\left(1-\frac{C^2}{36}\right)\theta^{6-\frac{3\left(24-C^2\right)}{2\left(36-C^2\right)}}\sum_{\tilde{k}}b_{\tilde{k}}\Bigg[\tilde{u}_1\exp\left(i\frac{\tilde{\lambda}}{3\hbar}\theta^3\right)\nonumber\\
		&+\tilde{u}_2\exp\left(-i\frac{\tilde{\lambda}}{3\hbar}\theta^3\right)\Bigg]\left[d_1\exp\left(\frac{\sqrt{\tilde{k}^2}}{\hbar}z\right)+d_2\exp\left(-\frac{\sqrt{\tilde{k}^2}}{\hbar}z\right)\right],\\
		\hbar^2\chi_{\tilde{k}}\diffp[2]{\varphi_{\tilde{k}}}{z}&\approx \sqrt{\frac{6\hbar}{\pi\tilde{\lambda}}}\tilde{k}^2\theta^{-\frac{3\left(24-C^2\right)}{2\left(36-C^2\right)}}\sum_{\tilde{k}}b_{\tilde{k}}\Bigg[\tilde{u}_1\exp\left(i\frac{\tilde{\lambda}}{3\hbar}\theta^3\right)\nonumber\\
		&+\tilde{u}_2\exp\left(-i\frac{\tilde{\lambda}}{3\hbar}\theta^3\right)\Bigg]\left[d_1\exp\left(\frac{\sqrt{\tilde{k}^2}}{\hbar}z\right)+d_2\exp\left(-\frac{\sqrt{\tilde{k}^2}}{\hbar}z\right)\right],\\
		\frac{C^2}{36}\hbar^2\theta\varphi_{\tilde{k}}\diffp{\chi_{\tilde{k}}}{\theta}&\approx \frac{C^2}{36}\sqrt{\frac{6\tilde{\lambda}\hbar^3}{\pi}}\theta^{3-\frac{3\left(24-C^2\right)}{2\left(36-C^2\right)}}\sum_{\tilde{k}}b_{\tilde{k}}\Bigg[\tilde{u}_1\exp\left(i\frac{\tilde{\lambda}}{3\hbar}\theta^3\right)\nonumber\\
		&-\tilde{u}_2\exp\left(-i\frac{\tilde{\lambda}}{3\hbar}\theta^3\right)\Bigg]\left[d_1\exp\left(\frac{\sqrt{\tilde{k}^2}}{\hbar}z\right)+d_2\exp\left(-\frac{\sqrt{\tilde{k}^2}}{\hbar}z\right)\right],\\
		\frac{\gamma-1}{6\lambda^2\gamma}\theta^6\chi_{\tilde{k}}\varphi_{\tilde{k}}&\approx\frac{\gamma-1}{6\lambda^2\gamma}\sqrt{\frac{6\hbar}{\pi\tilde{\lambda}}}\theta^{6-\frac{3\left(24-C^2\right)}{2\left(36-C^2\right)}}\sum_{\tilde{k}}b_{\tilde{k}}\Bigg[\tilde{u}_1\exp\left(i\frac{\tilde{\lambda}}{3\hbar}\theta^3\right)\nonumber\\
		&+\tilde{u}_2\exp\left(-i\frac{\tilde{\lambda}}{3\hbar}\theta^3\right)\Bigg]\left[d_1\exp\left(\frac{\sqrt{\tilde{k}^2}}{\hbar}z\right)+d_2\exp\left(-\frac{\sqrt{\tilde{k}^2}}{\hbar}z\right)\right].
	\end{align}
	\end{small}
	However, the terms that we have neglected behave asymptotically as
	\begin{small}
	\begin{align}
		\frac{C}{3}\hbar^2\theta\diffp{\chi_{\tilde{k}}}{\theta}\diffp{\varphi_{\tilde{k}}}{z}\approx&i\sqrt{\frac{6\tilde{\lambda}\tilde{k}^2}{\pi\hbar}}\frac{C\hbar}{3}\theta^{3-\frac{3\left(24-C^2\right)}{2\left(36-C^2\right)}}\sum_{\tilde{k}}b_{\tilde{k}}\Bigg[\tilde{u}_1\exp\left(i\frac{\tilde{\lambda}}{3\hbar}\theta^3\right)-\tilde{u}_2\exp\left(-i\frac{\tilde{\lambda}}{3\hbar}\theta^3\right)\Bigg]\nonumber\\
		&\times\left[d_1\exp\left(\frac{\sqrt{\tilde{k}^2}}{\hbar}z\right)-d_2\exp\left(-\frac{\sqrt{\tilde{k}^2}}{\hbar}z\right)\right].	
	\end{align}
	\end{small}
	We recall that the integration constants $d_1$ must be set to zero to fulfil the DW criterion at the BR when  $\tilde{k}^2$ positive.
	Thus, to obtain the compliance region of the performed approximation we compare the largest of the neglected terms with the smallest of the saved ones. This is the ratio 
	\begin{eqnarray}
		\epsilon=\left|\frac{	\frac{B}{3}\hbar^2\theta\partial_\theta{\chi_{\tilde{k}}}\partial_z{\varphi_{\tilde{k}}}}{\hbar^2\chi_{\tilde{k}}\partial^2_z{\varphi_{\tilde{k}}}}\right|\approx \frac{2}{3\tilde{k}}\frac{\gamma-2}{\gamma-1}\tilde{\lambda}\theta^3.
	\end{eqnarray}
	This ratio keeps below one if $\gamma-2$ is sufficiently small to compensate the increase of the variable $\theta$ towards the BR singularity. Therefore, for $\gamma\approx2$, this approximation is valid throughout the semiclassical regime towards the BR singularity, where $\theta$ increases but not sufficiently rapidly to compensate the small value of $\gamma-2$. In the expression (\ref{eq:BRfR}) for the metric $f(R)$ theory of gravity predicting the BR singularity, this would correspond to having a small parameter $A$ and $c_+=0$, since $\gamma_+$ diverge at $A\to0$. Consequently, this argument would favor small deviations from the $\Lambda$CDM model. It is worth noting that the values estimated for $A$ in reference \cite{CosmoConstraints} are not small enough to ensure $\epsilon\ll1$; however  smaller values for $A$ are compatible with the fits claimed in other references (see, for example, references \cite{Planck1,Planck2,Abbott:2018wog}). Thus, we consider this approximation to $\Psi$ valid for the appropriate $\gamma$ value. 
	
	Additionally note that in this section, we have enhanced the discussion on the viability of the performed approximations for $\Psi$ originally presented in reference \cite{BRfR}, where the authors have focused only on the particular case of $\tilde{k}=0$ when addressing the validity of the wave function there found.
	
	
	\section{Validity of the BO Approximation for $\Psi$\label{ap:BO}}
	During the application of the BO-type ansatz (\ref{eq:BOLR}) performed in Sections \ref{sec:mWDWLR} and \ref{sec:mWDWLSBR}, we have considered that  $\chi_{\tilde{ k}}(q,x)$ depends adiabatically on $x$. Therefore, we have neglected the contribution of some parts in their corresponding mWDW equations. This approach is valid as long as the corresponding solutions satisfy
	\begin{align}
		\hbar^2\varphi_{\tilde{ k}}\diffp[2]{\chi_{\tilde{ k}}}{x}, \ 2\hbar^2\diffp{\chi_{\tilde{ k}}}{ x}\diff{\varphi_{\tilde{ k}}}{x}\ll \hbar^2q^2\varphi_{\tilde{ k}}\diffp[2]{\chi_{\tilde{ k}}}{q},\ \hbar^2\chi_{\tilde{ k}}\diff[2]{\varphi_{\tilde{ k}}}{x},\ U(x)q^6\chi_{\tilde{ k}}\varphi_{\tilde{ k}}.
	\end{align}
	As a result of this approximation, the solutions for $\varphi_{\tilde{ k}}$ and $\chi_{\tilde{k}}$ obtained for the case of the LR abrupt event are presented in equations (\ref{eq:LR BO varphi x}) and (\ref{eq:LR BO chi qx}), respectively (see also \mbox{reference \cite{LRfR}}). Then, the terms we keep in (\ref{eq:LRBOmWDW}) read
	\begin{align}
		\hbar^2q^2\varphi_{\tilde{k}}\diffp[2]{\chi_{\tilde{k}}}{q}\approx&-U(x)q^6\chi_{\tilde{k}}\varphi_{\tilde{k}}\approx-\sqrt{\frac{6\hbar}{\pi}}U(x)^\frac34q^5\left[\tilde{u}_1\exp \left(i\frac{\sqrt{U(x)}}{3\hbar}q^3\right)\right.\nonumber\\
		&\left.+\tilde{u}_2\exp \left(-i\frac{\sqrt{U(x)}}{3\hbar}q^3\right)\right]\left[d_1\exp\left(\frac{\sqrt{\tilde{ k}^2}}{\hbar}x\right)\right.\left.+d_2\exp\left(-\frac{\sqrt{\tilde{ k}^2}}{\hbar}x\right)\right],\end{align}
	\begin{align}
		\hbar^2\chi_{\tilde{k}}\diff[2]{\varphi_{\tilde{k}}}{x}\approx&\sqrt{\frac{6\hbar}{\pi}}\frac{\tilde{k}^2}{U(x)^{\frac14}q}\left[\tilde{u}_1\exp \left(i\frac{\sqrt{U(x)}}{3\hbar}q^3\right)+\tilde{u}_2\exp \left(-i\frac{\sqrt{U(x)}}{3\hbar}q^3\right)\right]\nonumber\\
		&\times\left[d_1\exp\left(\frac{\sqrt{\tilde{ k}^2}}{\hbar}x\right)+d_2\exp\left(-\frac{\sqrt{\tilde{ k}^2}}{\hbar}x\right)\right].
	\end{align}
	However, the neglected terms behave asymptotically as
	\begin{small}
	\begin{align}
		\hbar^2\varphi_{\tilde{k}}\diffp[2]{\chi_{\tilde{k}}}{x}\approx&-\frac{1}{36}\sqrt{\frac{6\hbar}{\pi}}\frac{U'(x)^2}{U(x)^{\frac54}}q^5 \left[\tilde{u}_1\exp \left(i\frac{\sqrt{U(x)}}{3\hbar}q^3\right)+\tilde{u}_2\exp \left(-i\frac{\sqrt{U(x)}}{3\hbar}q^3\right)\right]\nonumber\\
		&\times\left[d_1\exp\left(\frac{\sqrt{\tilde{ k}^2}}{\hbar}x\right)+d_2\exp\left(-\frac{\sqrt{\tilde{ k}^2}}{\hbar}x\right)\right],\\
		2\hbar^2\diffp{\chi_{\tilde{k}}}{x}\diff{\varphi_{\tilde{k}}}{x}\approx&\frac{i}{3}\sqrt{\frac{6\hbar\tilde{ k}^2}{\pi}}\frac{U'(x)}{U(x)^{\frac34}}q^2\left[\tilde{u}_1\exp \left(i\frac{\sqrt{U(x)}}{3\hbar}q^3\right)-\tilde{u}_2\exp \left(-i\frac{\sqrt{U(x)}}{3\hbar}q^3\right)\right]\nonumber\\
		&\times\left[d_1\exp\left(\frac{\sqrt{\tilde{ k}^2}}{\hbar}x\right)-d_2\exp\left(-\frac{\sqrt{\tilde{ k}^2}}{\hbar}x\right)\right].
	\end{align}
	\end{small}
	Please note that for  $\tilde{ k}^2$ positive, the constants $d_1$ must be zero to have a vanishing wave function at the LR.
	Thus, the validity of the BO approximation can be verified comparing the largest of the neglected terms with the smallest of the saved ones. This is the ratio $\varepsilon$,
	\begin{align}\label{eqVal a/2}
		\varepsilon=\left|\frac{\hbar^2\varphi_{\tilde{ k}}\partial^2_x\chi_{\tilde{ k}}}{\hbar^2\chi_{\tilde{ k}}\partial^2_x\varphi_{\tilde{ k}}}\right|.
	\end{align}
	For the case of the LR abrupt event analyzed in Section \ref{sec:mWDWLR} that ratio reads
	\begin{eqnarray}
		\varepsilon\approx\frac{U'(x)^2}{U(x)}\frac{q^6}{36|\tilde{ k}^2|}.
	\end{eqnarray}
	Consequently, the approximation is valid as long as $\varepsilon\ll1$.
	To evaluate this condition, note~that
	\begin{align}
		\frac{U'(x)^2}{U(x)}\approx36\frac{\beta^2}{\lambda^2R_\star}e^{-2x}\Bigg[1+\frac{40}{3}\sqrt{\frac{3c_1}{f_{R_\star}}}\beta e^{-x}+\frac{832}{3}\frac{c_1\beta^2}{f_{R_\star}}e^{-2x}+\mathcal{O}\left(e^{-3x}\right)\Bigg],
	\end{align}
	when $\beta$ is observationally constrained, see  Table \ref{tab:Cosmoconstraints}. Thus, in the configuration space near the LR cosmic event, we have
	\begin{eqnarray}
		\varepsilon\approx\frac{\beta^2}{\lambda^2R_\star|\tilde{ k}^2|}e^{-2x}q^6,
	\end{eqnarray}
	where both $q$ and $x$ diverge.
	Finally, $\varepsilon\ll1$ near the LR  if $\beta$ is sufficiently small, i.e., for small value of $A$. Please note that this corresponds, in fact,  to the observationally preferred \mbox{situation \cite{CosmoConstraints}}. (We recall that $A$ is of order $10^{-28}\textup{ m}^{-1}$ when observationally constrained, see \mbox{Table \ref{tab:Cosmoconstraints}}). 
	Therefore, when the parameters of the theory are observationally constrained, the approximation is valid throughout the semiclassical regime towards the LR abrupt event; where the variables $q$ and $x$ increase but not sufficiently rapidly to compensate the small value of $\beta^2$. Hence, for the purpose of the present work, i.e., to analyze the fulfilment of the DW criterion in the configuration space close to the LR, this approximation is valid.
	
	On the other hand, for the verification of the validity of the BO approximation performed in Section \ref{sec:mWDWLSBR}, $U(x)$ must be exchanged for $W(x)$ in the preceding formulas. Therefore, considering the expression for $W(x)$ given in (\ref{eq:WLSBR}), the ratio $\varepsilon$ becomes
	\begin{eqnarray}
		\varepsilon\approx\frac{W'(x)^2}{W(x)}\frac{q^6}{36|\tilde{ k}^2|}\approx\frac{6c_1A^2}{\lambda^2R_\star f_{R_\star}|\tilde{k}^2|}e^{-4x}q^6.
	\end{eqnarray}
	Following the same line of reasoning as that presented before, this ratio keeps below 1 since the parameter $A$ for the LSBR is of order $10^{-54} \textup{ m}^{-2}$, see Table \ref{tab:Cosmoconstraints}. Therefore, the BO approximation applied in Section \ref{sec:mWDWLSBR} is valid throughout the semiclassical regime towards the LSBR abrupt event.

\end{paracol}
\printendnotes[custom]

\reftitle{References}



\begin{thebibliography}{999}

\bibitem{Riess}
A.~G.~Riess {\it et al.} [Supernova Search Team],
``Observational evidence from supernovae for an accelerating universe and a cosmological constant'',
Astron.\ J.\  {\bf 116} (1998) 1009,
[\href{https://arxiv.org/abs/astro-ph/9805201}{arXiv:astro-ph/9805201}].

\bibitem{Perlmutter}
S.~Perlmutter {\it et al.} [Supernova Cosmology Project Collaboration],
``Measurements of Omega and Lambda from 42 high redshift supernovae'',
Astrophys.\ J.\  {\bf 517} (1999) 565,
[\href{https://arxiv.org/abs/astro-ph/9812133}{arXiv:astro-ph/9812133}].


\bibitem{Weinberg:2000yb}
S.~Weinberg, ``The Cosmological constant problems'', [\href{https://arxiv.org/abs/astro-ph/0005265}{arXiv:astro-ph/0005265 [astro-ph]}].

\bibitem{Peebles:2002gy}
P.~J.~E.~Peebles and B.~Ratra, ``The Cosmological constant and dark energy'', Rev.\ Mod.\ Phys.\  {\bf 75} (2003) 559,
[\href{https://arxiv.org/abs/astro-ph/0207347}{arXiv:astro-ph/0207347[astro-ph]}].

\bibitem{Padmanabhan:2002ji}
T.~Padmanabhan, ``Cosmological constant: The Weight of the vacuum'', Phys.\ Rept.\  {\bf 380} (2003) 235,
[\href{https://arxiv.org/abs/hep-th/0212290}{arXiv:hep-th/0212290 [hep-th]}].

\bibitem{Sahnidmde}
V.~Sahni, ``Dark matter and dark energy'', Lect. Notes Phys. \textbf{653} (2004), 141-180,
[\href{https://arxiv.org/abs/astro-ph/0403324}{arXiv:astro-ph/0403324 [astro-ph]}].

\bibitem{Copeland:2006wr}
E.~J.~Copeland, M.~Sami and S.~Tsujikawa, ``Dynamics of dark energy'', Int.\ J.\ Mod.\ Phys.\ D {\bf 15} (2006) 1753,
[\href{https://arxiv.org/abs/hep-th/0603057}{arXiv:hep-th/0603057 [hep-th]}].

\bibitem{Velten:2014nra}
H.~E.~S.~Velten, R.~F.~vom Marttens and W.~Zimdahl, ``Aspects of the cosmological `coincidence problem' '', Eur. Phys. J. C \textbf{74} (2014) no.11, 3160,
[\href{https://arxiv.org/abs/1410.2509}{arXiv:1410.2509 [astro-ph.CO]}].

\bibitem{LambdaProblem}
R.~J.~Adler, B.~Casey and O.~C.~Jacob, ``Vacuum catastrophe: An elementary exposition of the cosmological constant problem'', Am. J. Phys. \textbf{63} (1995) 620-626.

\bibitem{Bull:2015stt}
P.~Bull, Y.~Akrami, J.~Adamek, T.~Baker, E.~Bellini, J.~Beltran Jimenez, E.~Bentivegna, S.~Camera, S.~Clesse and J.~H.~Davis, \textit{et al.}, ``Beyond $\Lambda$CDM: Problems, solutions, and the road ahead'', Phys. Dark Univ. \textbf{12} (2016) 56-99,
[\href{https://arxiv.org/abs/1512.05356}{arXiv:1512.05356 [astro-ph.CO]}].

\bibitem{Perivolaropoulos:2021jda}
L.~Perivolaropoulos and F.~Skara, ``Challenges for $\Lambda$CDM: An update'',
[\href{https://arxiv.org/abs/2105.05208}{arXiv:2105.05208 [astro-ph.CO]}].



\bibitem{PhantomDE}
R.~R.~Caldwell, ``A Phantom menace?'', Phys. Lett. B \textbf{545} (2002), 23-29,
[\href{https://arxiv.org/abs/astro-ph/9908168}{arXiv:astro-ph/9908168 [astro-ph]}].

\bibitem{Starobinsky:1999yw}
A.~A.~Starobinsky, ``Future and origin of our universe: Modern view'', Grav. Cosmol. \textbf{6} (2000), 157-163,
[\href{https://arxiv.org/abs/astro-ph/9912054}{arXiv:astro-ph/9912054 [astro-ph]}].


\bibitem{TachyonGibbons:2002md}
G.~W.~Gibbons, ``Cosmological evolution of the rolling tachyon'', Phys. Lett. B \textbf{537} (2002), 1-4,  
[\href{https://arxiv.org/abs/hep-th/0204008}{arXiv:hep-th/0204008 [hep-th]}].

\bibitem{TachyonPadmanabhan:2002cp}
T.~Padmanabhan, ``Accelerated expansion of the universe driven by tachyonic matter'', Phys. Rev. D \textbf{66} (2002), 021301,
[\href{https://arxiv.org/abs/hep-th/0204150}{arXiv:hep-th/0204150 [hep-th]}].


\bibitem{ChaplyginKamenshchik:2001cp}
A.~Y.~Kamenshchik, U.~Moschella and V.~Pasquier, ``An Alternative to quintessence'', Phys. Lett. B \textbf{511} (2001), 265-268,
[\href{https://arxiv.org/abs/gr-qc/0103004}{arXiv:gr-qc/0103004 [gr-qc]}].

\bibitem{ChaplyginBento:2002ps}
M.~C.~Bento, O.~Bertolami and A.~A.~Sen, ``Generalized Chaplygin gas, accelerated expansion and dark energy matter unification'', Phys. Rev. D \textbf{66} (2002), 043507,
[\href{https://arxiv.org/abs/gr-qc/0202064}{arXiv:gr-qc/0202064 [gr-qc]}].


\bibitem{HoloLi:2004rb}
M.~Li, ``A Model of holographic dark energy'', Phys. Lett. B \textbf{603} (2004), 1,
[\href{https://arxiv.org/abs/hep-th/0403127}{arXiv:hep-th/0403127 [hep-th]}].

\bibitem{Caldwell}
R.~R.~Caldwell, R.~Dave and P.~J.~Steinhardt, ``Cosmological imprint of an energy component with general equation of state'', Phys. Rev. Lett. \textbf{80} (1998), 1582-1585, 
[\href{https://arxiv.org/abs/astro-ph/9708069}{arXiv:astro-ph/9708069 [astro-ph]}].

\bibitem{Tsujikawa}
S.~Tsujikawa, ``Quintessence: A Review'', Class. Quant. Grav. \textbf{30} (2013), 214003,
[\href{https://arxiv.org/abs/1304.1961}{arXiv:1304.1961 [gr-qc]}].


\bibitem{kessenceChiba:1999ka}
T.~Chiba, T.~Okabe and M.~Yamaguchi, ``Kinetically driven quintessence'',
Phys. Rev. D \textbf{62} (2000), 023511,
[\href{https://arxiv.org/abs/astro-ph/9912463}{arXiv:astro-ph/9912463 [astro-ph]}].


\bibitem{Horndeski}
G.~W.~Horndeski, ``Second-order scalar-tensor field equations in a four-dimensional space'', Int. J. Mod. Phys. \textbf{10} (1974), 363-384.

\bibitem{Kase:2018aps}
R.~Kase and S.~Tsujikawa, ``Dark energy in Horndeski theories after GW170817: A review'', Int. J. Mod. Phys. D \textbf{28} (2019) no.05, 1942005,
[\href{https://arxiv.org/abs/1809.08735}{arXiv:1809.08735 [gr-qc]}].


\bibitem{GB1Dehghani:2004cf}
M.~H.~Dehghani, ``Accelerated expansion of the Universe in Gauss-Bonnet gravity'', Phys. Rev. D \textbf{70} (2004), 064009, 
[\href{https://arxiv.org/abs/hep-th/0404118}{arXiv:hep-th/0404118 [hep-th]}].

\bibitem{GB2Nojiri:2005jg}
S.~Nojiri and S.~D.~Odintsov, ``Modified Gauss-Bonnet theory as gravitational alternative for dark energy'', Phys. Lett. B \textbf{631} (2005), 1-6,
[\href{https://arxiv.org/abs/hep-th/0508049v2}{arXiv:hep-th/0508049 [hep-th]}].


\bibitem{DEinfRNojiri}
S.~Nojiri and S.~D.~Odintsov,
``Modified f(R) gravity consistent with realistic cosmology: From matter dominated epoch to dark energy universe'',
Phys. Rev. D \textbf{74} (2006), 086005,
[\href{https://arxiv.org/abs/hep-th/0608008}{arXiv:hep-th/0608008 [hep-th]}].

\bibitem{Nojiri:2004bi}
S.~Nojiri and S.~D.~Odintsov, ``Gravity assisted dark energy dominance and cosmic acceleration'', Phys.\ Lett.\ B {\bf 599} (2004), 137,
[\href{https://arxiv.org/abs/astro-ph/0403622}{astro-ph/0403622}].

\bibitem{Allemandi:2005qs}
G.~Allemandi, A.~Borowiec, M.~Francaviglia and S.~D.~Odintsov, ``Dark energy dominance and cosmic acceleration in first order formalism'', Phys.\ Rev.\ D {\bf 72} (2005), 063505,
[\href{https://arxiv.org/abs/gr-qc/0504057}{gr-qc/0504057}].

\bibitem{Capozziello:2011et}
S.~Capozziello and M.~De Laurentis, ``Extended Theories of Gravity'', Phys. Rept. \textbf{509} (2011), 167-321,
[\href{https://arxiv.org/abs/1108.6266}{arXiv:1108.6266 [gr-qc]}].


\bibitem{DEinfRTHarko:2011kv}
T.~Harko, F.~S.~N.~Lobo, S.~Nojiri and S.~D.~Odintsov, ``$f(R,T)$ gravity'',
Phys. Rev. D \textbf{84} (2011), 024020,
[\href{https://arxiv.org/abs/1104.2669}{arXiv:1104.2669 [gr-qc]}].


\bibitem{DEinfTBengochea:2008gz}
G.~R.~Bengochea and R.~Ferraro, ``Dark torsion as the cosmic speed-up'', Phys. Rev. D \textbf{79} (2009), 124019,
[\href{https://arxiv.org/abs/0812.1205}{arXiv:0812.1205 [astro-ph]}].


\bibitem{Jimenez:2019ovq}
J.~Beltr\'an Jim\'enez, L.~Heisenberg, T.~S.~Koivisto and S.~Pekar, ``Cosmology in $f(Q)$ geometry'', Phys. Rev. D \textbf{101} (2020) no.10, 103507,
[\href{https://arxiv.org/abs/1906.10027}{arXiv:1906.10027 [gr-qc]}].

\bibitem{Saridakis:2021lqd}
E.~N.~Saridakis \textit{et al.} [CANTATA], ``Modified Gravity and Cosmology: An Update by the CANTATA Network'',
[\href{https://arxiv.org/abs/2105.12582}{arXiv:2105.12582 [gr-qc]}].

\bibitem{Huterer:2017buf}
D.~Huterer and D.~L.~Shafer, ``Dark energy two decades after: Observables, probes, consistency tests'', Rept. Prog. Phys. \textbf{81} (2018) no.1, 016901,
[\href{https://arxiv.org/abs/1709.01091}{arXiv:1709.01091 [astro-ph.CO]}].

\bibitem{reviewDE}
K.~Bamba, S.~Capozziello, S.~Nojiri and S.~D.~Odintsov, ``Dark energy cosmology: the equivalent description via different theoretical models and cosmography tests'', Astrophys. Space Sci. \textbf{342} (2012), 155-228,
[\href{https://arxiv.org/abs/1205.3421v3}{arXiv:1205.3421 [gr-qc]}].

\bibitem{Motta:2021hvl}
V.~Motta, M.~A.~Garc\'\i{}a-Aspeitia, A.~Hern\'andez-Almada, J.~Maga\~na and T.~Verdugo, ``Taxonomy of Dark Energy Models'', Universe \textbf{7} (2021) no.6, 163,
[\href{https://arxiv.org/abs/2104.04642}{arXiv:2104.04642 [astro-ph.CO]}].


\bibitem{Planck1}
N.~Aghanim \textit{et al.} [Planck], ``Planck 2018 results. VI. Cosmological parameters'', Astron. Astrophys. \textbf{641} (2020), A6
[\href{https://arxiv.org/abs/1807.06209}{arXiv:1807.06209 [astro-ph.CO]}].

\bibitem{Planck2}  
P.~A.~R.~Ade {\it et al.} [Planck Collaboration], ``Planck 2015 results. XIII. Cosmological parameters'', Astron.\ Astrophys.\  {\bf 594} (2016) A13,  
[\href{https://arxiv.org/abs/1502.01589}{arXiv:1502.01589 [astro-ph.CO]}].

\bibitem{Abbott:2018wog}
T.~M.~C.~Abbott \textit{et al.} [DES], ``First Cosmology Results using Type Ia Supernovae from the Dark Energy Survey: Constraints on Cosmological Parameters'',
Astrophys. J. Lett. \textbf{872} (2019) no.2, L30,
[\href{https://arxiv.org/abs/1811.02374}{arXiv:1811.02374 [astro-ph.CO]}].

\bibitem{Lopez-Corredoira:2016pwg}
M.~Lopez-Corredoira, F.~Melia, E.~Lusso and G.~Risaliti, ``Cosmological test with the QSO Hubble diagram'', Int. J. Mod. Phys. D \textbf{25} (2016) no.05, 1650060,
[\href{https://arxiv.org/abs/1602.06743}{arXiv:1602.06743 [astro-ph.CO]}].

\bibitem{wCDM1}
K.~Vanderlinde \textit{et al.}, ``Galaxy Clusters Selected with the Sunyaev-Zel'dovich Effect from 2008 South Pole Telescope Observations'', Astrophys. J. \textbf{722} (2010) 1180, 
[\href{https://arxiv.org/abs/1003.0003}{arXiv:1003.0003 [astro-ph.CO]}].

\bibitem{wDCM2}
N.~Sehgal \textit{et al.}, ``The Atacama Cosmology Telescope: Cosmology from Galaxy Clusters Detected via the Sunyaev-Zel'dovich Effect'', Astrophys. J. \textbf{732} (2011) 44-45, 
[\href{https://arxiv.org/abs/1010.1025}{arXiv:1010.1025 [astro-ph.CO]}].

\bibitem{Addison:2013haa}
G.~E.~Addison, G.~Hinshaw and M.~Halpern, ``Cosmological constraints from baryon acoustic oscillations and clustering of large-scale structure'', Mon. Not. Roy. Astron. Soc. \textbf{436} (2013), 1674-1683,
[\href{https://arxiv.org/abs/1304.6984}{arXiv:1304.6984 [astro-ph.CO]}].

\bibitem{Santos:2015nua}
M.~Vargas dos Santos, R.~R.~R.~Reis and I.~Waga, ``Constraining the cosmic deceleration-acceleration transition with type Ia supernova, BAO/CMB and H(z) data'', JCAP \textbf{02} (2016), 066,
[\href{https://arxiv.org/abs/1505.03814}{arXiv:1505.03814 [astro-ph.CO]}].

\bibitem{Bonilla:2017ygx}
A.~Bonilla and J.~E.~Castillo, ``Constraints On Dark Energy Models From Galaxy Clusters and Gravitational Lensing Data'', Universe \textbf{4} (2018) no.1, 21,
[\href{https://arxiv.org/abs/1711.09291}{arXiv:1711.09291 [astro-ph.CO]}].



\bibitem{RisalitiLusso}
G.~Risaliti and E.~Lusso, ``Cosmological constraints from the Hubble diagram of quasars at high redshifts'', Nature Astron. \textbf{3} (2019) no.3, 272-277,
[\href{https://arxiv.org/abs/1811.02590}{arXiv:1811.02590 [astro-ph.CO]}].

\bibitem{EDVH01}
E.~Di Valentino, A.~Melchiorri, E.~V.~Linder and J.~Silk, ``Constraining Dark Energy Dynamics in Extended Parameter Space'', Phys. Rev. D \textbf{96} (2017) no.2, 023523,
[\href{https://arxiv.org/abs/1704.00762}{arXiv:1704.00762 [astro-ph.CO]}].

\bibitem{EDVH02}
E.~Di Valentino, E.~V.~Linder and A.~Melchiorri, ``Vacuum phase transition solves the $H_0$ tension'', Phys. Rev. D \textbf{97} (2018) no.4, 043528,
[\href{https://arxiv.org/abs/1710.02153}{arXiv:1710.02153 [astro-ph.CO]}].


\bibitem{BouhmadiLopez:2004me}
M.~Bouhmadi-L\'opez and J.~A.~Jimenez Madrid, ``Escaping the big rip?'', JCAP \textbf{05} (2005), 005,
[\href{https://arxiv.org/abs/astro-ph/0404540}{arXiv:astro-ph/0404540 [astro-ph]}].


\bibitem{BR}
R.~R.~Caldwell, M.~Kamionkowski and N.~N.~Weinberg, ``Phantom Energy: Dark Energy with $w<\ensuremath{-}1$ Causes a Cosmic Doomsday'', Phys. Rev. Lett. \textbf{91} (2003), 071301
[\href{https://arxiv.org/abs/astro-ph/0302506}{arXiv:astro-ph/0302506 [astro-ph]}].

\bibitem{LR}
P.~H.~Frampton, K.~J.~Ludwick and R.~J.~Scherrer, ``The Little Rip'', Phys.\ Rev.\ D {\textbf{ 84}} (2011) 063003,
[\href{https://arxiv.org/abs/1106.4996}{arXiv:1106.4996 [astro-ph.CO]}].

\bibitem{EOSalpha2table}
H.~\u{S}tefan\u{c}i\'c, ``Expansion around the vacuum equation of state - Sudden future singularities and asymptotic behavior'',	Phys. Rev. D \textbf{71} (2005) 084024,
[\href{https://arxiv.org/abs/astro-ph/0411630v3}{arXiv:astro-ph/0411630 [astro-ph]]}.

\bibitem{BouhmadiLopez:2005gk}
M.~Bouhmadi-L\'opez, ``Phantom-like behaviour in dilatonic brane-world scenario with induced gravity'', Nucl. Phys. B \textbf{797} (2008), 78-92,
[\href{https://arxiv.org/abs/astro-ph/0512124}{arXiv:astro-ph/0512124 [astro-ph]}].

\bibitem{LR2}
P.~H.~Frampton, K.~J.~Ludwick, S.~Nojiri, S.~D.~Odintsov and R.~J.~Scherrer, ``Models for Little Rip Dark Energy'', Phys.\ Lett.\ B {\textbf{708}} (2012) 204,
[\href{https://arxiv.org/abs/1108.0067v2}{arXiv:1108.0067 [hep-th]}].

\bibitem{LSBR}
M.~Bouhmadi-L\'opez, A.~Errahmani, P.~Mart\'in-Moruno, T.~Ouali and Y.~Tavakoli,  ``The little sibling of the big rip singularity'',
Int.\ J.\ Mod.\ Phys.\ D {\bf 24} (2015) no.10,  1550078,
[\href{https://arxiv.org/abs/1407.2446}{arXiv:1407.2446 [gr-qc]}].

\bibitem{CosmoConstraints2}
I.~Albarran, M.~Bouhmadi-L\'opez and J.~Morais, ``Cosmological perturbations in an effective and genuinely phantom dark energy Universe'', Phys. Dark Univ. \textbf{16} (2017), 94-108,
[\href{https://arxiv.org/abs/1611.00392}{arXiv:1611.00392 [astro-ph.CO]}].

\bibitem{CosmoConstraints}A.~Bouali, I.~Albarran, M.~Bouhmadi-L\'opez and T.~Ouali, ``Cosmological constraints of phantom dark energy models'', 	Phys. Dark Univ. \textbf{26} (2019), 100391,
[\href{https://arxiv.org/abs/1905.07304}{arXiv:1905.07304 [astro-ph.CO]}].

\bibitem{BF1BouhmadiLopez:2006fu}
M.~Bouhmadi-L\'opez, P.~F.~Gonz\'alez-D\'iaz and P.~Mart\'in-Moruno, ``Worse than a big rip?'', Phys. Lett. B \textbf{659} (2008), 1-5, 
[\href{https://arxiv.org/abs/gr-qc/0612135}{arXiv:gr-qc/0612135 [gr-qc]}].

\bibitem{BF2BouhmadiLopez:2007qb}
M.~Bouhmadi-L\'opez, P.~F.~Gonz\'alez-D\'iaz and P.~Mart\'in-Moruno, ``On the generalised Chaplygin gas: Worse than a big rip or quieter than a sudden singularity?'', Int. J. Mod. Phys. D \textbf{17} (2008), 2269-2290,
[\href{https://arxiv.org/abs/0707.2390}{arXiv:0707.2390 [gr-qc]}].


\bibitem{sudden1}
J.~D.~Barrow, ``Sudden future singularities'', Class. Quant. Grav. \textbf{21} (2004), L79-L82,
[\href{https://arxiv.org/abs/gr-qc/0403084}{arXiv:gr-qc/0403084 [gr-qc]}].

\bibitem{sudden2}
K.~Lake, ``Sudden future singularities in FLRW cosmologies'',
Class. Quant. Grav. \textbf{21} (2004), L129,
[\href{https://arxiv.org/abs/gr-qc/0407107}{arXiv:gr-qc/0407107 [gr-qc]}].

\bibitem{sudden3}
J.~D.~Barrow, ``More general sudden singularities'', Class. Quant. Grav. \textbf{21} (2004), 5619-5622,
[\href{https://arxiv.org/abs/gr-qc/0409062}{arXiv:gr-qc/0409062 [gr-qc]}].


\bibitem{Dabrowski:2014fha}
M.~P.~D\c{a}browski, ``Are singularities the limits of cosmology?'',
[\href{https://arxiv.org/abs/1407.4851}{arXiv:1407.4851 [gr-qc]}].

\bibitem{EOSalpha3}
S.~Nojiri, S.~D.~Odintsov and S.~Tsujikawa, ``Properties of singularities in (phantom) dark energy universe'', Phys. Rev. D \textbf{71} (2005) 063004,
[\href{https://arxiv.org/abs/hep-th/0501025}{arXiv:hep-th/0501025 [hep-th]}].

\bibitem{SingularityClasification}
M.~Bouhmadi-L\'opez, C.~Kiefer and P.~Mart\'in-Moruno, ``Phantom singularities and their quantum fate: general relativity and beyond\textemdash{}a CANTATA COST action topic'',
Gen. Rel. Grav. \textbf{51} (2019) no.10, 135,
[\href{https://arxiv.org/abs/1904.01836}{arXiv:1904.01836 [gr-qc]}].

\bibitem{Dabrowski:2006dd}
M.~P.~D\c{a}browski, C.~Kiefer and B.~Sandh\"ofer, ``Quantum phantom cosmology'',
Phys. Rev. D \textbf{74} (2006), 044022,
[\href{https://arxiv.org/abs/hep-th/0605229}{arXiv:hep-th/0605229 [hep-th]}].

\bibitem{Kamenshchik:2007zj}
A.~Kamenshchik, C.~Kiefer and B.~Sandh\"ofer, ``Quantum cosmology with big-brake singularity'', Phys. Rev. D \textbf{76} (2007), 064032,
[\href{https://arxiv.org/abs/0705.1688}{arXiv:0705.1688 [gr-qc]}].

\bibitem{EOSalpha1}
S.~Nojiri and S.~D.~Odintsov, ``The Final state and thermodynamics of dark energy universe'',	Phys. Rev. D \textbf{70} (2004) 103522,
[\href{https://arxiv.org/abs/hep-th/0408170}{arXiv:hep-th/0408170 [hep-th]}].

\bibitem{Elizalde:2004mq}
E.~Elizalde, S.~Nojiri and S.~D.~Odintsov, ``Late-time cosmology in (phantom) scalar-tensor theory: Dark energy and the cosmic speed-up'', Phys. Rev. D \textbf{70} (2004), 043539,
[\href{https://arxiv.org/abs/hep-th/0405034}{arXiv:hep-th/0405034 [hep-th]}].

\bibitem{Nojiri:2004ip}
S.~Nojiri and S.~D.~Odintsov, ``Quantum escape of sudden future singularity'', Phys. Lett. B \textbf{595} (2004), 1-8,
[\href{https://arxiv.org/abs/hep-th/0405078}{arXiv:hep-th/0405078 [hep-th]}].


\bibitem{BouhmadiLopez:2009pu}
M.~Bouhmadi-L\'opez, C.~Kiefer, B.~Sandh\"ofer and P.~Vargas Moniz, ``On the quantum fate of singularities in a dark-energy dominated universe'',
Phys. Rev. D \textbf{79} (2009), 124035,
[\href{https://arxiv.org/abs/0905.2421}{arXiv:0905.2421 [gr-qc]}].

\bibitem{Albarran:2015cda}
I.~Albarran, M.~Bouhmadi-L\'opez, F.~Cabral and P.~Mart\'\i{}n-Moruno, ``The quantum realm of the ''Little Sibling'' of the Big Rip singularity',
JCAP \textbf{11} (2015), 044,
[\href{https://arxiv.org/abs/1509.07398}{arXiv:1509.07398 [gr-qc]}].

\bibitem{GRLR}
I.~Albarran, M.~Bouhmadi-L\'opez, C.~Kiefer, J.~Marto and P.~Vargas Moniz, ``Classical and quantum cosmology of the little rip abrupt event'', Phys. Rev. D \textbf{94} (2016) no.6, 063536,
[\href{https://arxiv.org/abs/1604.08365}{arXiv:1604.08365 [gr-qc]}].

\bibitem{DeWitt}
B.~S.~DeWitt, ``Quantum Theory of Gravity. I. The Canonical Theory'',  Phys. Rev. \textbf{160} (1967) 1113.

\bibitem{Vilenkin}
A.~Vilenkin, ``Classical and Quantum Cosmology of the Starobinsky Inflationary Model'', Phys. Rev. D \textbf{32} (1985) 2511.

\bibitem{Fernandez-Jambrina:2014sga}
L.~Fern\'andez-Jambrina, ``Grand Rip and Grand Bang/Crunch cosmological singularities'', Phys. Rev. D \textbf{90} (2014), 064014,
[\href{https://arxiv.org/abs/1408.6997}{arXiv:1408.6997 [gr-qc]}].

\bibitem{Fernandez-Jambrina:2016clh}
L.~Fern\'andez-Jambrina, ``Initial directional singularity in inflationary models'', Phys. Rev. D \textbf{94} (2016) no.2, 024049,
[\href{https://arxiv.org/abs/1606.07600}{arXiv:1606.07600 [gr-qc]}].

\bibitem{Ruzmaikina}
T.~V.~Ruzma\u{\i}kina, A.~A.~Ruzma\u{\i}kin, ``Quadratic corrections to the Lagrangian density of the gravitational field and the singularity", Sov. Phys. JETP 30 (1970) 372.


\bibitem{Bouali:2021upl}
A.~Bouali, I.~Albarran, M.~Bouhmadi-L\'opez, A.~Errahmani and T.~Ouali, ``Cosmological constraints of interacting phantom dark energy models'',
[\href{https://arxiv.org/abs/2103.13432}{arXiv:2103.13432 [astro-ph.CO]}].


\bibitem{ReconNojiri99pag}
S.~Nojiri and S.~D.~Odintsov, ``Unified cosmic history in modified gravity: from F(R) theory to Lorentz non-invariant models'', Phys. Rept. \textbf{505} (2011), 59-144,
[\href{https://arxiv.org/abs/1011.0544}{arXiv:1011.0544 [gr-qc]}].


\bibitem{ReconCapozziello}
S.~Capozziello, V.~F.~Cardone and A.~Troisi, ``Reconciling dark energy models with $f(R)$ theories'', Phys.\ Rev.\ D {\bf 71} (2005) 043503, 
[\href{https://arxiv.org/abs/astro-ph/0501426}{arXiv: astro-ph/0501426}].

\bibitem{ReconfRgravity}
\'A.~de la Cruz-Dombriz and A.~Dobado, ``A $f(R)$ gravity without cosmological constant'', Phys. Rev. D \textbf{74} (2006) 087501, [\href{https://arxiv.org/abs/gr-qc/0607118}{arXiv:grqc/0607118}].

\bibitem{ReconMGravity}
S.~Nojiri and S.~D.~Odintsov, ``Modified gravity and its reconstruction from the universe expansion history'', J. Phys. Conf. Ser. \textbf{66} (2007) 012005, [\href{https://arxiv.org/abs/hep-th/0611071}{arXiv:hep-th/0611071}].

\bibitem{ReconNojiri}
S.~Nojiri, S.~D.~Odintsov, A.~Toporensky and P.~Tretyakov, ``Reconstruction and deceleration-acceleration transitions in modified gravity'',
Gen. Rel. Grav. \textbf{42} (2010), 1997-2008,
[\href{https://arxiv.org/abs/0912.2488}{arXiv:0912.2488 [hep-th]}].

\bibitem{ReconDunsby}
P.~K.~S.~Dunsby, E.~Elizalde, R.~Goswami, S.~Odintsov and D.~S.~G\'omez, ``On the LCDM Universe in f(R) gravity'',
Phys. Rev. D \textbf{82} (2010), 023519
[\href{https://arxiv.org/abs/1005.2205}{arXiv:1005.2205 [gr-qc]}].

\bibitem{ReconCarloni}
S.~Carloni, R.~Goswami and P.~K.~S.~Dunsby, ``A new approach to reconstruction methods in $f(R)$ gravity'', Clas. Quant. Grav. \textbf{29} (2012) 135012, [\href{https://arxiv.org/abs/1005.1840}{arXiv:1005.1840 [gr-qc]}].

\bibitem{ReconRef1} 
J.~Morais, M.~Bouhmadi-L\'opez and S.~Capozziello, ``Can $f(R)$ gravity contribute to (dark) radiation?'', 
JCAP {\bf 1509} (2015) no.09,  041, 
[\href{https://arxiv.org/abs/1507.02623}{arXiv:1507.02623 [gr-qc]}].


\bibitem{LRinfHG}
A.~N.~Makarenko, V.~V.~Obukhov and I.~V.~Kirnos, ``From Big to Little Rip in modified F(R,G) gravity'', Astrophys.\ Space Sci.\  {\textbf{343}} (2013) 481,
[\href{https://arxiv.org/abs/1201.4742}{arXiv:1201.4742 [gr-qc]}].

\bibitem{BRfR} 
A.~Alonso-Serrano, M.~Bouhmadi-L\'opez and P.~Mart\'in-Moruno,  ``$f(R)$ quantum cosmology: avoiding the Big Rip'', Phys.  Rev.  D {\bf 98} (2018) no. 10, 104004,
[\href{https://arxiv.org/abs/1802.03290}{arXiv:1802.03290 [gr-qc]}].

\bibitem{LSBRfR}
T.~Borislavov Vasilev, M.~Bouhmadi-L\'opez and P.~Mart\'in-Moruno, ``Classical and quantum fate of the little sibling of the big rip in $f(R)$ cosmology'',
Phys.\ Rev.\ D {\bf 100} (2019) no. 8,  084016,
[\href{https://arxiv.org/abs/1907.13081}{arXiv:1907.13081 [gr-qc]}].

\bibitem{LRfR}
T.~Borislavov Vasilev, M.~Bouhmadi-L\'opez and P.~Mart\'\i{}n-Moruno, ``The little rip in classical and quantum $f(R)$ cosmology'', Phys.\ Rev.\ D \textbf{103} (2021) no. 12, 124049,
[\href{https://arxiv.org/abs/2103.12786}{arXiv:2103.12786 [gr-qc]}].

\bibitem{libroFunciones} M.~Abramowitz and I.~A.~Stegun, \textit{Handbook of Mathematical Functions}, Dover Publications, New York (1972), ISBN 978-0486612720.


\bibitem{Cassini}
B.~Bertotti, L.~Iess and P.~Tortora, ``A test of general relativity using radio links with the Cassini spacecraft'', Nature \textbf{425} (2003) 374–376. 

\bibitem{Capozziello:2005bu}
S.~Capozziello and A.~Troisi,p ``PPN-limit of fourth order gravity inspired by scalar-tensor gravity'', Phys. Rev. D \textbf{72} (2005), 044022,
[\href{https://arxiv.org/abs/astro-ph/0507545}{arXiv:astro-ph/0507545 [astro-ph]}].

\bibitem{Chiba:2006jp}
T.~Chiba, T.~L.~Smith and A.~L.~Erickcek, ``Solar System constraints to general f(R) gravity'', Phys. Rev. D \textbf{75} (2007), 124014,
[\href{https://arxiv.org/abs/astro-ph/0611867}{arXiv:astro-ph/0611867 [astro-ph]}].

\bibitem{Capozziello:2007ms}
S.~Capozziello, A.~Stabile and A.~Troisi, ``The Newtonian Limit of f(R) gravity'',
Phys. Rev. D \textbf{76} (2007), 104019,
[\href{https://arxiv.org/abs/0708.0723}{arXiv:0708.0723 [gr-qc]}].

\bibitem{Capozziello:2009zz}
S.~Capozziello, E.~Piedipalumbo, C.~Rubano and P.~Scudellaro, ``Testing an exact f(R)-gravity model at Galactic and local scales'', Astron. Astrophys. \textbf{505} (2009), 21-28,
[\href{https://arxiv.org/abs/0906.5430}{arXiv:0906.5430 [gr-qc]}].

\bibitem{ODwyer:2013vfo}
M.~O'Dwyer, S.~E.~Joras and I.~Waga, ``$\gamma$ gravity: Steepness control'', Phys. Rev. D \textbf{88} (2013) no.6, 063520,
[\href{https://arxiv.org/abs/1305.4654}{arXiv:1305.4654 [astro-ph.CO]}].

\bibitem{Wang:2019svo}
J.~Y.~Wang, C.~J.~Feng, X.~H.~Zhai and X.~Z.~Li, ``Solar System Tests of a New Class of $f(z)$ Theory'', Int. J. Mod. Phys. D \textbf{29} (2020) no.08, 2050060,
[\href{https://arxiv.org/abs/1905.12212}{arXiv:1905.12212 [astro-ph.CO]}].

\bibitem{Sawicki:2007tf}
I.~Sawicki and W.~Hu, ``Stability of Cosmological Solution in f(R) Models of Gravity'', Phys. Rev. D \textbf{75} (2007), 127502,
[\href{https://arxiv.org/abs/astro-ph/0702278v2}{arXiv:astro-ph/0702278 [astro-ph]}].

\bibitem{Hu:2007nk}
W.~Hu and I.~Sawicki, ``Models of f(R) Cosmic Acceleration that Evade Solar-System Tests'', Phys. Rev. D \textbf{76} (2007), 064004,
[\href{https://arxiv.org/abs/0705.1158v1}{arXiv:0705.1158 [astro-ph]}].

\bibitem{Roshan:2011kz}
M.~Roshan and F.~Shojai, ``Notes on the post-Newtonian limit of massive Brans-Dicke theory'', Class. Quant. Grav. \textbf{28} (2011), 145012,
[\href{https://arxiv.org/abs/1106.1264}{arXiv:1106.1264 [gr-qc]}].

\bibitem{Guo:2013fda}
J.~Q.~Guo, ``Solar system tests of f(R) gravity'',
Int. J. Mod. Phys. D \textbf{23} (2014) no.4, 1450036,
[\href{https://arxiv.org/abs/1306.1853}{arXiv:1306.1853 [astro-ph.CO]}].

\bibitem{Naik:2018mtx}
A.~P.~Naik, E.~Puchwein, A.~C.~Davis and C.~Arnold, ``Imprints of Chameleon f(R) Gravity on Galaxy Rotation Curves'', Mon. Not. Roy. Astron. Soc. \textbf{480} (2018) no.4, 5211-5225,
[\href{https://arxiv.org/abs/1805.12221}{arXiv:1805.12221 [astro-ph.CO]}].

\bibitem{Negrelli:2020yps}
C.~Negrelli, L.~Kraiselburd, S.~J.~Landau and M.~Salgado, ``Solar System tests and chameleon effect in $f(R)$ gravity'',
Phys. Rev. D \textbf{101} (2020) no.6, 064005,
[\href{https://arxiv.org/abs/2002.12073}{arXiv:2002.12073 [gr-qc]}].



\bibitem{Klauslibro}
C.~Kiefer, {\em Quantum gravity}, 3rd edition. Oxford University
Press, Oxford (2012), ISBN 978-0199212521.

\bibitem{KlausBarbara} 
C.~Kiefer and B.~Sandh\"oefer, ``Quantum Cosmology'', [\href{https://arxiv.org/abs/0804.0672}{arXiv:0804.0672 [gr-qc]}].


\bibitem{Bojowald:2006da}
M.~Bojowald, ``Loop quantum cosmology'', Living Rev. Rel. \textbf{8} (2005), 11,
[\href{https://arxiv.org/abs/gr-qc/0601085}{arXiv:gr-qc/0601085 [gr-qc]}].

\bibitem{Ashtekar:2011ni}
A.~Ashtekar and P.~Singh, ``Loop Quantum Cosmology: A Status Report'', Class. Quant. Grav. \textbf{28} (2011), 213001,
[\href{https://arxiv.org/abs/1108.0893}{arXiv:1108.0893 [gr-qc]}].


\bibitem{CDT1}
J.~Ambj{\o}rn and R.~Loll, ``Nonperturbative Lorentzian quantum gravity, causality and topology change'', Nucl. Phys. B \textbf{536} (1998), 407-434, 
[\href{https://arxiv.org/abs/hep-th/9805108}{arXiv:hep-th/9805108 [hep-th]}].

\bibitem{CDT2}
J.~Ambj{\o}rn, J.~Jurkiewicz and R.~Loll, ``A Nonperturbative Lorentzian path integral for gravity'', Phys. Rev. Lett. \textbf{85} (2000), 924-927,
[\href{https://arxiv.org/abs/hep-th/0002050}{arXiv:hep-th/0002050 [hep-th]}].

\bibitem{CDT3}
J.~Ambj{\o}rn, J.~Jurkiewicz and R.~Loll, ``Dynamically triangulating Lorentzian quantum gravity'', Nucl. Phys. B \textbf{610} (2001), 347-382,
[\href{https://arxiv.org/abs/hep-th/0105267}{arXiv:hep-th/0105267 [hep-th]}].



\bibitem{Kuchar:1989tj}
K.~V.~Kuchar and M.~P.~Ryan, ``Is minisuperspace quantization valid?: Taub in mixmaster'', Phys. Rev. D \textbf{40} (1989), 3982-3996.

\bibitem{Wheeler}
J.~A.~Wheeler, ``On the Nature of quantum geometrodynamics'', Annals Phys.\  {\bf 2} (1957) 604.

\bibitem{Bouhmadi-Lopez:2016dcf}
M.~Bouhmadi-L\'opez and C.~Y.~Chen, ``Towards the Quantization of Eddington-inspired-Born-Infeld Theory'', JCAP \textbf{11} (2016), 023,
[\href{https://arxiv.org/abs/1609.00700}{arXiv:1609.00700 [gr-qc]}].

\bibitem{Bouhmadi-Lopez:2018tel}
M.~Bouhmadi-L\'opez, C.~Y.~Chen and P.~Chen, ``On the Consistency of the Wheeler-DeWitt Equation in the Quantized Eddington-inspired Born-Infeld Gravity'', JCAP \textbf{12} (2018), 032,
[\href{https://arxiv.org/abs/1810.10918}{arXiv:1810.10918 [gr-qc]}].

\bibitem{Albarran:2018mpg}
I.~Albarran, M.~Bouhmadi-L\'opez, C.~Y.~Chen and P.~Chen, ``Quantum cosmology of Eddington-Born\textendash{}Infeld gravity fed by a scalar field: The big rip case'', Phys. Dark Univ. \textbf{23} (2019), 100255,
[\href{https://arxiv.org/abs/1811.05041}{arXiv:1811.05041 [gr-qc]}].



\bibitem{EvsJframe1}
P.~W.~Higgs, ``Quadratic lagrangians and general relativity'', Nuovo Cim \textbf{11} (1959) 816 . 

\bibitem{EvsJrframe2}
G.~V.~Bicknell, ``Non-viability of gravitational theory based on a
quadratic lagrangian'', J.\ Phys.\ A: Math.\ Nucl.\ Gen.\ \textbf{7} (1974) 1061. 

\bibitem{EvsJrframe3}
B.~Whitt, ``Fourth-order gravity as general relativity plus matter'', Phys.\ Lett.\ B \textbf{145} (1984) 176. 

\bibitem{EvsJrframe4}	 
J.~D.~Barrow and S.~Cotsakis, ``Inflation and the Conformal Structure of Higher Order Gravity Theories'', Phys. Lett. B \textbf{214} (1988) 515.


\bibitem{noLSBR2}
I.~Albarran, M.~Bouhmadi-L\'opez, F.~Cabral and P.~Mart\'\i{}n-Moruno, ``The Avoidance of the Little Sibling of the Big Rip Abrupt Event by a Quantum Approach'', Galaxies \textbf{6} (2018) 21. 



\bibitem{EiBI_Albarran:2017swy}
I.~Albarran, M.~Bouhmadi-L\'opez, C.~Y.~Chen and P.~Chen,
``Doomsdays in a modified theory of gravity: A classical and a quantum approach'', Phys. Lett. B \textbf{772} (2017), 814-818,
[\href{https://arxiv.org/abs/1703.09263}{arXiv:1703.09263 [gr-qc]}].


\bibitem{ellipticWDW}
C.~Kiefer, ``Non-minimally coupled scalar fields and the initial value problem in quantum gravity'', Phys. Lett. B \textbf{225} (1989) 227.


\bibitem{BOoriginal}
M.~Born and J.~R.~Oppenheimer, ``On quantum theory of molecules'', Ann.\ Physik\ \textbf{389} (1927) 457.


\bibitem{ClausBO1}
C.~Kiefer, ``Continuous measurement of mini-superspace variables by higher multipoles'', Class. Quantum Grav. \textbf{4} (1987) 1369.

\bibitem{BOenWDW}
R.~Brout, G.~Horwitz and D.~Weil, ``On the onset of time and temperature in cosmology'', Phys.\ Lett.\ B \textbf{192} (1987) 318.

\bibitem{ClausBO2}
C.~Kiefer, ``Wave packets in minisuperspace'',	Phys.\ Rev.\ D \textbf{38} (1988) 1761.
	
\end{thebibliography}


\end{document}